\documentclass[pdflatex,sn-mathphys-num]{sn-jnl}

\usepackage{graphicx}%
\usepackage{multirow}%
\usepackage{amsmath,amssymb,amsfonts}%
\usepackage{amsthm}%
\usepackage{mathrsfs}%
\usepackage[title]{appendix}%
\usepackage{xcolor}%
\usepackage{textcomp}%
\usepackage{manyfoot}%
\usepackage{booktabs}%
\usepackage{algorithm}%
\usepackage{algorithmicx}%
\usepackage{algpseudocode}%
\usepackage{listings}%
\usepackage[T1]{fontenc}
\usepackage{ulem}
\newcommand{\updates}[1]{#1}

\theoremstyle{thmstyleone}%

\theoremstyle{thmstyletwo}%

\theoremstyle{thmstylethree}%

\renewcommand{\orcidlogo}{%
        \includegraphics[width=10pt]{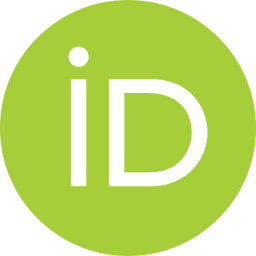}%
}

\def\i{{\rm i}}
\def\f{{\rm f}}
\def\Q{{\rm Q}}
\def\e{{\rm e}}

\begin{document}

\title[Article Title]{Kibble-Zurek Scaling and Spatial Statistics in Quenched Binary Bose Superfluids}

%\author[1]{\fnm{Subhadeep}
%\sur{Patra}
%\orcidlink{0009-0009-2041-8539}
%}\email{d23222@students.iitmandi.ac.in}

%\author*[1]{\fnm{Arko} \sur{Roy} \orcidlink{0000-0003-4459-2880}}\email{arko@iitmandi.ac.in}

%\author[2]{\fnm{Seong-Ho} \sur{Shinn} \orcidlink{0000-0002-2041-5292}}\email{seongho.shin@uni.lu}

%\author[2,3]{\fnm{Adolfo} \sur{del Campo} \orcidlink{0000-0003-2219-2851}}\email{adolfo.delcampo@uni.lu}

%\author[4]{\fnm{Mithun} \sur{Thudiyangal} \orcidlink{0000-0003-4341-6439}}\email{mithun.thudiyangal@christuniversity.in}

%\affil[1]{School of Physical Sciences, Indian Institute of Technology Mandi, Mandi-175005 (H.P.), India}

%\affil[2]{Department  of  Physics  and  Materials  Science,  University  of  Luxembourg,  L-1511  Luxembourg,  Luxembourg}

%\affil[3]{Donostia International Physics Center,  E-20018 San Sebasti\'an, Spain}

%\affil[4]{Center for Quantum Technologies and Complex Systems (CQTCS), Christ University, Bengaluru, Karnataka 560029, India}

\author[1]{\fnm{Subhadeep}\sur{Patra}{\href{https://orcid.org/0009-0009-2041-8539}{\orcidlogo}}}\email{d23222@students.iitmandi.ac.in}

\author*[1]{\fnm{Arko} \sur{Roy}{\href{https://orcid.org/0000-0003-4459-2880}{\orcidlogo}}}\email{arko@iitmandi.ac.in}

\author[2]{\fnm{Seong-Ho} \sur{Shinn}{\href{https://orcid.org/0000-0002-2041-5292}{\orcidlogo}}}\email{seongho.shin@uni.lu}
\author[2,3]{\fnm{Adolfo} \sur{del Campo}{\href{https://orcid.org/0000-0003-2219-2851}{\orcidlogo}}}\email{adolfo.delcampo@uni.lu}
\author[4]{\fnm{Mithun} \sur{Thudiyangal}{\href{https://orcid.org/0000-0003-4341-6439}{\orcidlogo}}}\email{mithun.thudiyangal@christuniversity.in}

\affil[1]{School of Physical Sciences, Indian Institute of Technology Mandi, Mandi-175005 (H.P.), India}
\affil[2]{Department  of  Physics  and  Materials  Science,  University  of  Luxembourg,  L-1511  Luxembourg,  Luxembourg}
\affil[3]{Donostia International Physics Center,  E-20018 San Sebasti\'an, Spain}
\affil[4]{Center for Quantum Technologies and Complex Systems (CQTCS), Christ University, Bengaluru, Karnataka 560029, India}

% (SHS comment) Editors requested to rewrite the abstract. 
\abstract{\updates{The emergence of order from an initially uncorrelated state across a phase transition is a central problem in quantum many-body physics, particularly in multicomponent systems where interactions between components lead to rich nonequilibrium dynamics. While defect formation is known to follow universal scaling laws, prior studies have focused mainly on defect density, leaving their spatial organization largely unexplored. Here we show that gradually tuning the chemical potential in a two-dimensional binary Bose gas drives condensation into either a miscible or immiscible phase. In the immiscible regime, domains form whose number, size, and boundary length obey Kibble–Zurek (KZ) scaling and evolve self-similarly. In the miscible regime, vortices emerge with KZ scaling. In both cases, the spatial distribution of vortices and domains is well described by a Poisson point process with KZ-determined density. These results reveal universal features of far-from-equilibrium dynamics and provide a framework to characterize stochastic geometry in multicomponent quantum systems.}}

% \abstract{We study how gradually changing the chemical potential causes a two-dimensional binary Bose gas to condense from vacuum to finite density, resulting in either a mixed (miscible) or separated (immiscible) state depending on interaction strengths. In the immiscible case, random domains form, and their number, boundary length, and average size at the point of equilibration follow universal Kibble-Zurek (KZ) scaling with the cooling rate. 
% These patterns continue to evolve in a self-similar way while maintaining KZ scaling. In the miscible regime, instead of domains, vortices appear, following the KZ scaling. The spatial distribution of domains and vortices is described by a Poisson point process with the KZ density. These findings highlight robust, universal features of how binary superfluids behave far from equilibrium, extending beyond the KZ theory.}

% (SHS comment) Keywords are not supported and requested to be removed by the editors
%\keywords{Kibble-Zurek mechanism, ultra-cold quantum gases, superfluidity, quantum phase transitions, non-equilibrium statistical mechanics, non-linear pattern formation, phase-separation, universal dynamics}

\maketitle

\section{Introduction}\label{sec1}

The dynamics of spontaneous symmetry breaking through phase transition has been studied in various fields, including cosmology \cite{Kibble_1976,kibble_1980,Morikawa_1995}, condensed matter physics \cite{Buerle1996,Ruutu1996,Monaco_2009},  ultracold atom systems  \cite{Zurek1985,Ueda10,StamperKurn13,DZ14}, and their use for quantum simulation \cite{Keesling2019,Chepiga2021,Ebadi2021}.
Ultracold atomic gases provide an ideal testbed for studying such dynamics in a controlled and tunable manner. Furthermore, the ability to drive phase transitions in these systems enables direct probing of universal scaling laws, such as those predicted by the
Kibble-Zurek mechanism (KZM)~\cite{Kibble_1976,Zurek1985,Zurek_1996,DZ14}. Ultracold gases thus provide a unique platform for exploring defect formation and critical dynamics in a non-equilibrium setting. The cooling dynamics of a thermal gas leading to the formation of a scalar Bose-Einstein condensate (BEC) of a single component is characterized by $U(1)$ symmetry breaking and can lead to the formation of solitons or vortices, depending on the dimensionality \cite{Weiler2008,Zurek_2009,Lamporesi2013,Navon_2015,Clark_2016,Feng2018,Shin19,Goo21,Thudiyangal_2024}. A richer arena is offered by spinor gases \cite{StamperKurn13,Anquez_2016,Maximilian_2018} and atomic mixtures \cite{Baroni2024}. For instance, in the case of Bose-Bose mixtures~\cite{Ho_1996,Hofmann_2014,Trippenbach_2000}, two interacting superfluid components evolve under external tuning parameters such as chemical potential or interaction strength. In these mixtures, phase transitions give rise to rich defect structures, including quantized vortices and spatial domains, whose formation reflects the interplay of microscopic interactions and collective critical dynamics. 

The KZM theory predicts that the density of topological defects resulting from the critical dynamics scales as a universal power law of the quench time in which the transition is crossed, with the power-law exponent being determined by the critical exponents and the dimensionality of the system. To date, KZM has been studied in both classical and quantum phase transitions \cite{DZ14} and tested in various setups, including trapped bosonic gas~\cite{Donadello_2016, Scherer_2007, delCampo2011, delCampo2013,Lamporesi2013,Goo21}, trapped ions \cite{Ulm13,Pyka13,Cui19}, superconducting vortex systems~\cite{Reichhardt_2022, Maegochi_2022}, and Josephson junctions~\cite{Monaco2006}. 

In this work, we investigate the formation of a Bose-Bose mixture driven by a linear quench of the chemical potential, resulting in the growth of either a miscible or an immiscible phase, depending on the inter- and intra-species interactions. 
This system provides the opportunity to study the formation of domains and vortices, as well as their universal dynamics, in a single platform.  
Domain formation has been explored in various ultracold systems, including spin-1 spin-orbit coupled BECs~\cite{Wu_2017}, binary BECs with internal coupling~\cite{Sabbatini_2011}, and confined systems in elongated harmonic traps~\cite{Sabbatini_2012} and ring-shaped quasi-1D potentials~\cite{Swisłock_2013}.
Unlike coherently
coupled spinor condensates where magnetization gradients develop across
domain walls, here the two components are coupled only through density–density interactions,  
and spontaneous phase
separation in the immiscible regime occurs purely via interaction energy
minimization.

% (SHS comment) brief summary of the approach, major results, and conclusions of this manuscript should be at the final paragraph of the introduction
\updates{
Using the stochastic (projected) Gross-Pitaevskii equation (SPGPE) in two dimensions~\cite{Ota_2018, Proukakis_2008} together with the quench protocol described in the Methods, we focus on the symmetry-breaking dynamics following condensation from vacuum. 
First, in the immiscible phase, we show the number of domains, domain areas, domain wall lengths, at a characteristic equilibration time $t_{\rm eq}$ to obey universal power-law behavior with respect to the quench time $\tau_{\rm Q}$. The early-time collapse of domain wall length curves indicates the self-similar dynamics. 
Second, in the miscible phase, we demonstrate vortex formation and verify Kibble-Zurek (KZ) scaling of the vortex numbers. 
Finally, going beyond defect counting, we studied the universal spatial statistics of vortices in the miscible phase and mean domain positions in the immiscible phase, showing that their distributions are well described by Poisson point processes (PPP), uncovering universal nonequilibrium dynamics that lie beyond the scope of the KZM. 
These predictions are amenable to direct experimental verification using atomic mixtures in uniform traps, and can be applied to other fields such as the circulation statistics or quantum turbulence.}

\section{Results \updates{and Discussions}}\label{sec3}

\subsection{Condensate norm evolution and characterization of the equilibration time}
\label{sec3-1}

% (SHS comment) As editors requested to move the "Methods" after the conclusions, I moved ``Stochastic formalism" and moved Figure 1 here. Position should be placed in a better way...
% (SHS comment) Editors requested to write brief titles in all figures.
\updates{
\begin{figure}[t]
    \centering
    \includegraphics[width=0.7\linewidth]{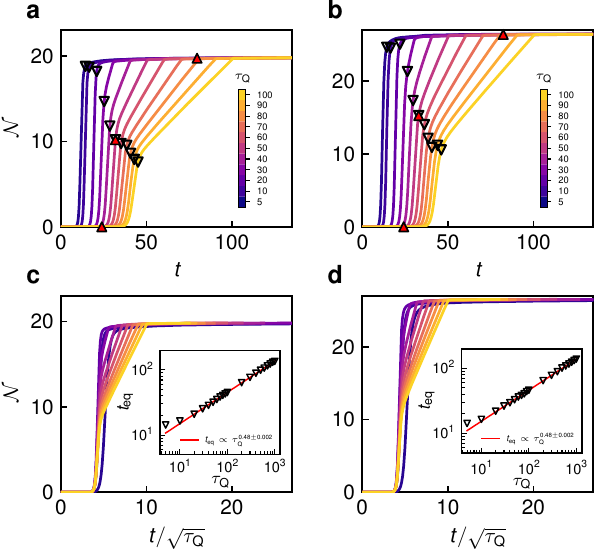}
    \caption{
    \updates{Scaled number density of atoms for various quench times.}
    Norm $\mathcal{N}(t)$ 
    that represents the scaled number density of atoms is plotted 
     \updates{as a function of time $t$} during growth into (a) immiscible and (b) miscible phases for quench times $
    %\tau_Q
    %(SHS comment) Editors requested to use Roman for subscripts and superscripts to be italic ONLY IF they are variables...
    \updates{\tau_{\Q}}
    = [5, 10, 20, \dots, 100]$, arranged left to right for a single noise realization. The equilibration time $
    %t_{\text{eq}}
    t_{\rm eq}
    $ (black triangles) scales as 
    $
    %t_{\text{eq}}
    t_{\rm eq}
    \propto 
    %\tau_Q
    \updates{\tau_{\Q}}
    ^{0.48 \pm 0.002}$
    for $g_{12} > g$ [inset of (c)] and 
    $
    %t_{\text{eq}} 
    t_{\rm eq}
    \propto 
    %\tau_Q
    \updates{\tau_{\Q}}
    ^{0.48 \pm 0.002}$
     for $g_{12} < g$ [inset of (d)], shown by red 
     %lines. 
     % (SHS comment) Editors explicitly requested to define every symbol in every Figure, even if they are defined in the main text.
     \updates{lines, where $g$ and $g_{12}$ are the intra- and inter-species interaction strengths, respectively.}
     (c, d) The early exponential growth ($t < 
     %t_{\text{eq}}
     t_{\rm eq}
     $) collapses when time is scaled by $
    %\tau_Q
    \updates{\tau_{\Q}}
     ^{1/2}$, except for $
     %\tau_Q
    \updates{\tau_{\Q}}
      = 5, 10$. Filled red markers on $
      %\tau_Q
    \updates{\tau_{\Q}}
      = 50$ curves in (a, b) indicate times \updates{at which} 
     %profiles in Figs.~\ref{density-growth} and \ref{mis-density-growth}.
     % (SHS comment) Editors requested not to refer to other figures before they have been referenced!
     \updates{the density profiles are presented later.}
     }
    \label{number-growth}
\end{figure}
}

With ${\mathcal A}$ being the area of the box, we monitor the growth of the norm $\mathcal{N}(t) = (1/\mathcal A){\sum_{j=1,2}\iint |\psi_j({\bf x},t)|^2d{\bf x}}$ of the order parameter corresponding to the symmetry broken phases illustrated in Fig.~\ref{number-growth} 
for different quench time 
$
%\tau_Q
\updates{\tau_{\Q}}
$. Note that the final scaled number density of atoms represented by $\mathcal{N}$ shown in Fig.~\ref{number-growth}, is consistent with 
%$\mu_f / g$. 
\updates{$\mu_{\f} / g$ where $\mu_{\f}$ is the final chemical potential after the quench and $g$ is the intra-species interaction strength (see the Methods).}
Details of the 
%SPGPE formalism, 
\updates{SPGPE formalism~\cite{Ota_2018, Proukakis_2008},}
along with a brief discussion of the numerical parameters, are provided in the 
%supplemental material \cite{SM}.
% (SHS comment) It seems that the editors do not like this \cite{SM}, saying that no information appears to have been provided, even though we clearly wrote ``supplemental material url will be inserted by publisher''...
% (SHS comment) General citations to the Supplementary Information should be avoided
\updates{Supplementary Methods.}
The onset of spontaneous symmetry breaking is signaled by a sharp increase in the condensate norm illustrated in Figs. \ref{number-growth}(a) and (b), 
accompanied by the emergence of protodomains.
% , as seen in Figs.~\ref{density-growth}(a) and \ref{mis-density-growth}(a) for the immiscible as well as miscible phases. 
In the early freeze-out stage (\( t \ll t_{\rm eq} \)), 
the condensate is ill-defined and lacks coherent structure 
\cite{chesler_2015}. A physically meaningful timescale for analyzing the KZ scaling is the equilibration time 
\( t_{\rm eq} \), which marks the crossover from exponential to adiabatic linear growth shown in Fig.~\ref{number-growth}. The procedure for extracting \( t_{\rm eq} \) is detailed 
%in \cite{SM}. 
\updates{in the Supplementary Note 1.}
At \( t_{\rm eq} \), which is proportional to the KZ freeze-out time, isolated protodomains begin to coalesce, signaling the onset of global order and symmetry breaking~\cite{chesler_2015}. It follows the KZ scaling similar to that of the freeze-out time, such that \( 
%t_{\text{eq}} 
t_{\rm eq} 
\propto \left( \tau_0 
%\tau_Q
\updates{\tau_{\Q}}
^{z\nu} \right)^{1/(1+z\nu)} \), with \( z\nu = 1 \), consistent with mean-field exponents \( z = 2 \), \( \nu = 1/2 \), and $\tau_0$ being a system dependent constant~\cite{Zurek_2009, Campo2021, Ma_2025, Thudiyangal_2024, MassimoCritical2001}.
Based on our numerical analysis, 
we extract 
$t_{\rm eq} = (4.97 \pm 0.072) 
%\tau_Q
\updates{\tau_{\Q}}
^{0.48 \pm 0.002}$
for the growth to the immiscible phase characterized by 
%$g_{12} > g$, 
\updates{$g_{12} > g$ with $g_{12}$ being the inter-species interaction strength,}
and 
$t_{\rm eq} = (5.13 \pm 0.064) 
%\tau_Q
\updates{\tau_{\Q}}
^{0.48 \pm 0.002}$
for the growth to the miscible phase characterized by $g_{12} < g$. Note that in our system the critical chemical potential $
%\mu_c
\updates{\mu_{\rm c}}
$, marking the symmetry breaking transition, is not crossed exactly at $t=0$. While this could introduce corrections to the KZM freeze-out time, we find their effect negligible, with the scaling dominated by the standard KZ power law; 
%see \cite{SM} 
\updates{see the Supplementary Note 1}
for further details.
The numerically obtained exponents match well with the KZM predictions. Deviations in the fast-quench regime (\( 
%\tau_Q
\updates{\tau_{\Q} }
< 20 \)) indicate a breakdown of KZ scaling, while the collapse of rescaled growth curves in Figs.~\ref{number-growth}(c) and (d) confirms the slow quench regime. These benchmarks aid to validate the KZM prediction for topological defect density \cite{DZ14, Ma_2025, chesler_2015, Ruiz_2020}: \( \rho \propto \left( \tau_0 / 
%\tau_Q 
\updates{\tau_{\Q}}
\right)^{(D - d)\nu / (1 + z\nu)} \), where \( D \) is the system dimension and \( d \) is the defect dimension (e.g., \( d = 0 \) for point defects). 
In general, thermal vortices may affect the KZ scaling on defect density. However, using the results in \cite{Giorgetti2007}, the number of thermal vortices 
per total number of bosons in the condensate 
is expected to be $\exp(- 6.7 \times 10^6)$ for the parameters we used. Therefore, 
since total number of bosons in the condensate is at most $\mathcal{N}(t) \mathcal{A} \approx 10^4$, 
the effect of the 
%BKT 
Berezinskii–Kosterlitz–Thouless (BKT)
transition on the KZ scaling is negligible, as in the quasi-two-dimensional Bose gas experiment \cite{Chomaz2015}.
\begin{figure}[!htbp]
    \centering
    \includegraphics[width=0.7\linewidth]{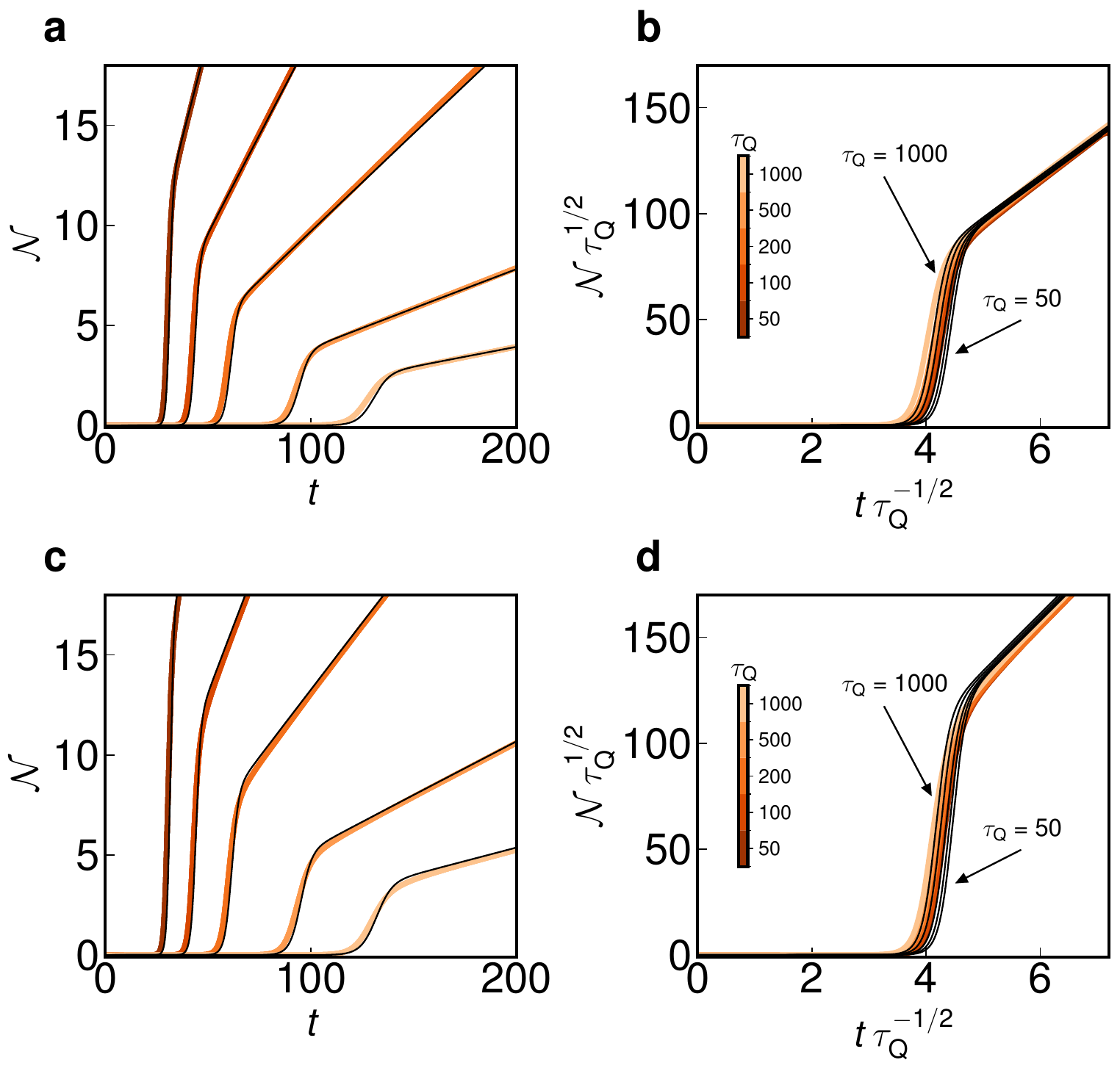}
    \caption{
    \updates{Testing the analytical prediction of the condensate norm.}
    %Analytical prediction of the condensate norm evolution $\mathcal{N}(t)$ in both phases.
(a) Time evolution of the condensate norm during growth into the immiscible phase for quench times
$
%\tau_Q
\updates{\tau_{\Q} }
= 50, 100, 200, 500,$ and $1000$. Thin black curves show the analytical solution of 
\updates{the condensate norm evolution $\mathcal{N}(t)$ as a function of time $t$ in}
Eq.~\ref{norm_analytic}, in excellent agreement with the numerical results (thick colored curves).
(b) Rescaled condensate norm $\mathcal{N}
%\tau_Q
\updates{\tau_{\Q}}
^{1/2}$ plotted as a function of the rescaled time
$t
%\tau_Q
\updates{\tau_{\Q}}
^{-1/2}$, demonstrating collapse of the growth curves and universal dynamics.
(c) Same as (a), but for growth into the miscible phase, where the analytical predictions again
closely reproduce the numerical evolution.
(d) Rescaled condensate norm $\mathcal{N}
%\tau_Q
\updates{\tau_{\Q}}
^{1/2}$ versus $t
%\tau_Q
\updates{\tau_{\Q}}
^{-1/2}$ for the miscible
phase, revealing the same universal growth behavior across different quench times $
%\tau_Q
\updates{\tau_{\Q}}
$.}
    \label{norm_analytic_fig}
\end{figure}

We further aim to capture the evolution of the condensate norm $\mathcal{N}(t)$ analytically. To this end, we consider the growth of the complex order parameter $\psi_j(\mathbf{x},t)$ governed by the SPGPE in Eq.~\ref{eq:sgpe}. By neglecting spatial dependence and stochastic noise, following \updates{Suzuki \textit{et al.}}~\cite{Suzuki_2025}, the stochastic partial differential equation can be reduced to an effective time-dependent ordinary differential equation, given by
\begin{equation}
    (1 - 
    %i
    % Editors requested to use Roman in e, i, and \pi... I defined "\i" in order to simply i in Roman.
    \updates{\i}
    \gamma)(g |\psi_j(t)|^2 \psi_j(t) + g_{12} |\psi_{3-j}(t)|^2 \psi_j(t) - \mu(t) \psi_j(t)) = 
    %i 
    \updates{\i} 
    \dot{\psi_j}(t)
    , 
    \label{temporal_ode}
\end{equation}
\updates{where $\gamma$ is the dissipation parameter and $\mu \left( t \right)$ is the chemical potential (see the Methods for details).}
Using Madelung transformation, Eq.~\ref{temporal_ode} becomes
\begin{equation}
    %\dot{n}_T
    \updates{\dot{n}_{\rm T}}
    = 2\gamma \mu(t)
    %n_T 
    \updates{n_{\rm T}}
    - \gamma (g + g_{12})(
    %n_T
    \updates{n_{\rm T}}
    ^2 + 
    %n_M
    \updates{n_{\rm M}}
    ^2) - \gamma (g - g_{12})
    %n_M
    \updates{n_{\rm M}}
    ^2.
\end{equation}
Here, the total density is $
%n_T 
\updates{n_{\rm T}}
= n_1 + n_2$ and the population imbalance is $
%n_M 
\updates{n_{\rm M}}
= n_1 - n_2$, with $n_1 = |\psi_1|^2$ and $n_2 = |\psi_2|^2$. During the growth stage, population imbalance remains negligible, allowing us to set $
%n_M 
\updates{n_{\rm M}}
= 0$. The evolution equation then reduces to
\begin{equation}
    %\dot{n}_T
    \updates{\dot{n}_{\rm T}}
    - 2\gamma \mu(t)
    %n_T 
    \updates{n_{\rm T}}
    = - \gamma (g + g_{12}) 
    %n_T
    \updates{n_{\rm T}}
    ^2,
    \label{bernoulli_ode}
\end{equation}
which admits an exact solution for the linear quench protocol
$\mu(t) = 
%\mu_i
\updates{\mu_{\i}}
+ (t/
%\tau_Q
\updates{\tau_{\Q}}
)(
%\mu_f 
\updates{\mu_{\f}}
- 
%\mu_i
\updates{\mu_{\i}}
)$
%. 
\updates{where $\mu_{\i}$ is the initial chemical potential before the quench.}
The resulting analytical expression for the condensate norm reads 
\begin{equation}
    \mathcal{N}(t) = 
    %n_T
    \updates{n_{\rm T}}
    (t) = \frac{\mathcal{N}(0) 
    %e
    % (SHS comment) Editors explicitly requested to write e in Roman... I introduced "\e" command to simplify it
    \updates{\e}
    ^{F(t)}}{1 + \mathcal{N}(0) KM(t)} ,
    \label{norm_analytic}
\end{equation}
where $\mathcal{N}(0)$ is the initial number density scaled by the simulation box area at $t = 0$ and $\mathrm{erfi}
(x)
= 2\int_0^x
%e
% (SHS comment) Editors explicitly requested to write e in Roman... I introduced "\e" command to simplify it
\updates{\e}
^{x'^2}dx'/\sqrt{
%\pi
\updates{\rm \pi}
}$. Here, $K$, $F(t)$, and $M(t)$ are defined as :
\begin{eqnarray*}
    K &=& \gamma(g + g_{12}), \\
    F(t) &=& 2\gamma 
    %\mu_i 
    \updates{\mu_{\i}}
    t + \frac{\gamma (
    %\mu_f
    \updates{\mu_{\f}}    
    - 
    %\mu_i
    \updates{\mu_{\i}}    
    )}{
    %\tau_Q
    \updates{\tau_{\Q}}
    } t^2, 
\end{eqnarray*}
and 
\begin{equation*}
    M(t) = \frac{\sqrt{
    %\pi
    \updates{\rm \pi} 
    }}{2}\sqrt{\frac{
    %\tau_Q
    \updates{\tau_{\Q}}
    }{\gamma (
    %\mu_f
    \updates{\mu_{\f}}
    - 
    %\mu_i
    \updates{\mu_{\i}} 
    )}}\,
    %e
    % (SHS comment) Editors explicitly requested to write e in Roman... I introduced "\e" command to simplify it
    \updates{\e}
    ^{-\frac{\gamma 
    %\tau_Q
    \updates{\tau_{\Q}}
    %\mu_i
    \updates{\mu_{\i}}
    ^2}{(
    %\mu_f
    \updates{\mu_{\f}}
    - 
    %\mu_i
    \updates{\mu_{\i}}
    )}}\mathrm{erfi}\bigg( \sqrt{\frac{\gamma (
    %\mu_f
    \updates{\mu_{\f}}
    - 
    %\mu_i
    \updates{\mu_{\i}}
    )}{
    %\tau_Q
    \updates{\tau_{\Q}}
    }}t + \sqrt{\frac{\gamma 
    %\tau_Q
    \updates{\tau_{\Q}}
    }{(
    %\mu_f
    \updates{\mu_{\f}}
    - 
    %\mu_i
    \updates{\mu_{\i}}
    )}} 
    %\mu_i 
    \updates{\mu_{\i}}
    \bigg),
\end{equation*}
respectively. 
The analytical prediction in Eq.~\ref{norm_analytic} quantitatively reproduces the numerical evolution of the condensate norm during growth into the immiscible phase, as shown in Fig.~\ref{norm_analytic_fig}(a) by the thin black curves for quench times $
%\tau_Q
\updates{\tau_{\Q} }
= 50, 100, 200, 500,$ and $1000$, demonstrating that the early-time condensate growth is well captured by the reduced description.
Further, we consider the KZ scaling of the condensate norm. For a periodic system, neglecting thermal noise term, $
%n_T
\updates{n_{\rm T}}
$ at the equilibration time is proportional to the chemical potential at that time as the system adiabatically follows the change of the chemical potential according to the KZM, giving $
%n_T
\updates{n_{\rm T}}
(t_{\rm eq}) \propto 
%\tau_Q
\updates{\tau_{\Q}}
^{-1 / \left( 1 + z \nu \right)}$ and thus $
%n_T
\updates{n_{\rm T}}
(t) \sim 
%\tau_Q
\updates{\tau_{\Q}}
^{-1 / \left( 1 + z \nu \right)} f( t / t_{\rm eq} ) = 
%\tau_Q
\updates{\tau_{\Q}}
^{- 1 / \left( 1 + z \nu \right)} f(t 
%\tau_Q
\updates{\tau_{\Q}}
^{-z \nu  /\left( 1 + z \nu \right)})$ for $t \geq t_{\rm eq}$ where $f(x)$ is some arbitrary function depending on $x$. This is consistent with the results in \cite{Suzuki_2025} as $z \nu = 1$ in our case, leading to 
$\mathcal{N}(t) \sim 
%\tau_Q
\updates{\tau_{\Q}}
^{-1/2}f(t
%\tau_Q
\updates{\tau_{\Q}}
^{-1/2})$ 
and revealing
a similar growth dynamics for different $
%\tau_Q
\updates{\tau_{\Q}}
$ shown in the Fig.~\ref{norm_analytic_fig}(b). We further analyze the norm evolution analytically in the miscible regime, as shown in Fig.~\ref{norm_analytic_fig}(c), where the analytical solution again captures the numerical growth of the condensate. In Fig.~\ref{norm_analytic_fig}(d), we present the appropriately scaled condensate norm, analogous to the immiscible case, and observe a comparable collapse of the curves for different quench times, indicating the same universal growth dynamics.

\subsection{Phase separated regime and self similar dynamics}\label{sec3-2}
% (SHS comment) Editors requested to write a brief title, and explain what color bars mean!
\begin{figure*}[!htbp]
    \centering
    \includegraphics[width=0.95\linewidth]{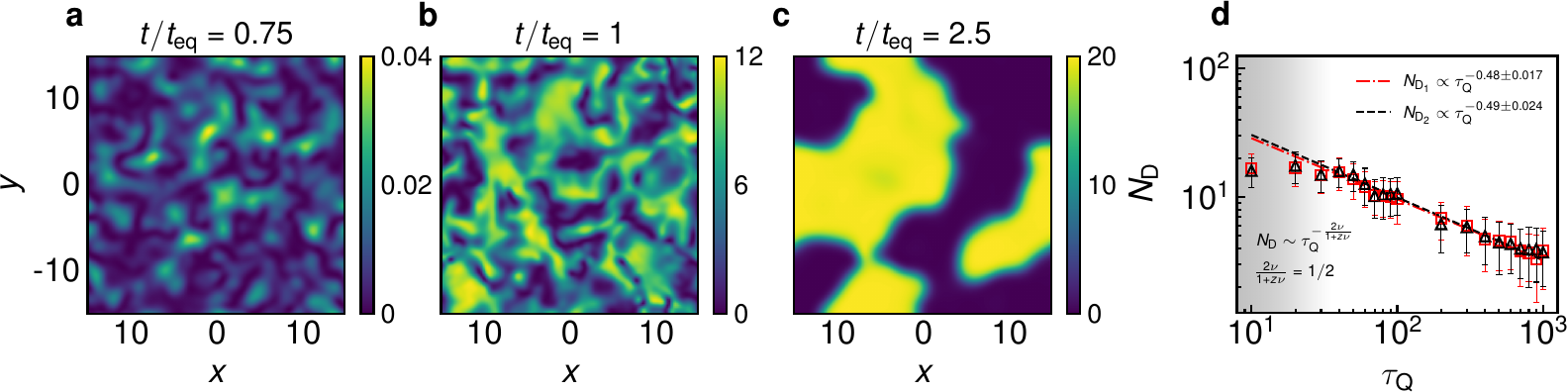}
    \caption{Condensate density evolution %\(|\psi_1|^2\)
    \updates{of the component 1 of the binary Bose-Bose mixture in the immiscible phase.}
    %\(\tau_Q = 50\).
 Panels (a)–(c) show domain formation and evolution %in the immiscible phase 
 \updates{as a function of time $t$ scaled by the equilibration time $t_{\rm eq}$ at the quench time $\tau_{\rm Q} = 50$} 
 , corresponding to red markers in Fig.~\ref{number-growth}(a). 
 \updates{Color bars represent the norm of the condensate order parameter.}
Panel (d) shows the domain number \(
%N_D
\updates{N_{\rm D}}
\) as a function of \(
%\tau_Q
\updates{\tau_{\Q}}
\) on a log scale, revealing 
%KZM-like
\updates{the Kibble-Zurek mechanism (KZM)}
scaling: 
\(
%N_{D_1} 
\updates{N_{\rm D_1}}
\propto 
%\tau_Q
\updates{\tau_{\Q}}
^{-0.48 \pm 0.017}\), \(
%N_{D_2} 
\updates{N_{\rm D_2}}
\propto 
%\tau_Q
\updates{\tau_{\Q}}
^{-0.49 \pm 0.024}\). Error bars show standard deviations across 100 noise realizations. Grey shaded area in (d) denotes deviation from universal KZM behavior, indicating a smooth crossover between fast and slow quenches.}
    \label{density-growth}
\end{figure*}
Domain formation is a characteristic feature of the immiscible phase, where the two condensate components spontaneously separate in space following the quench, as shown in Fig.~\ref{density-growth}(a)–(c). 

\begin{figure}[!htbp]
    \centering
    \includegraphics[width=0.7\linewidth]{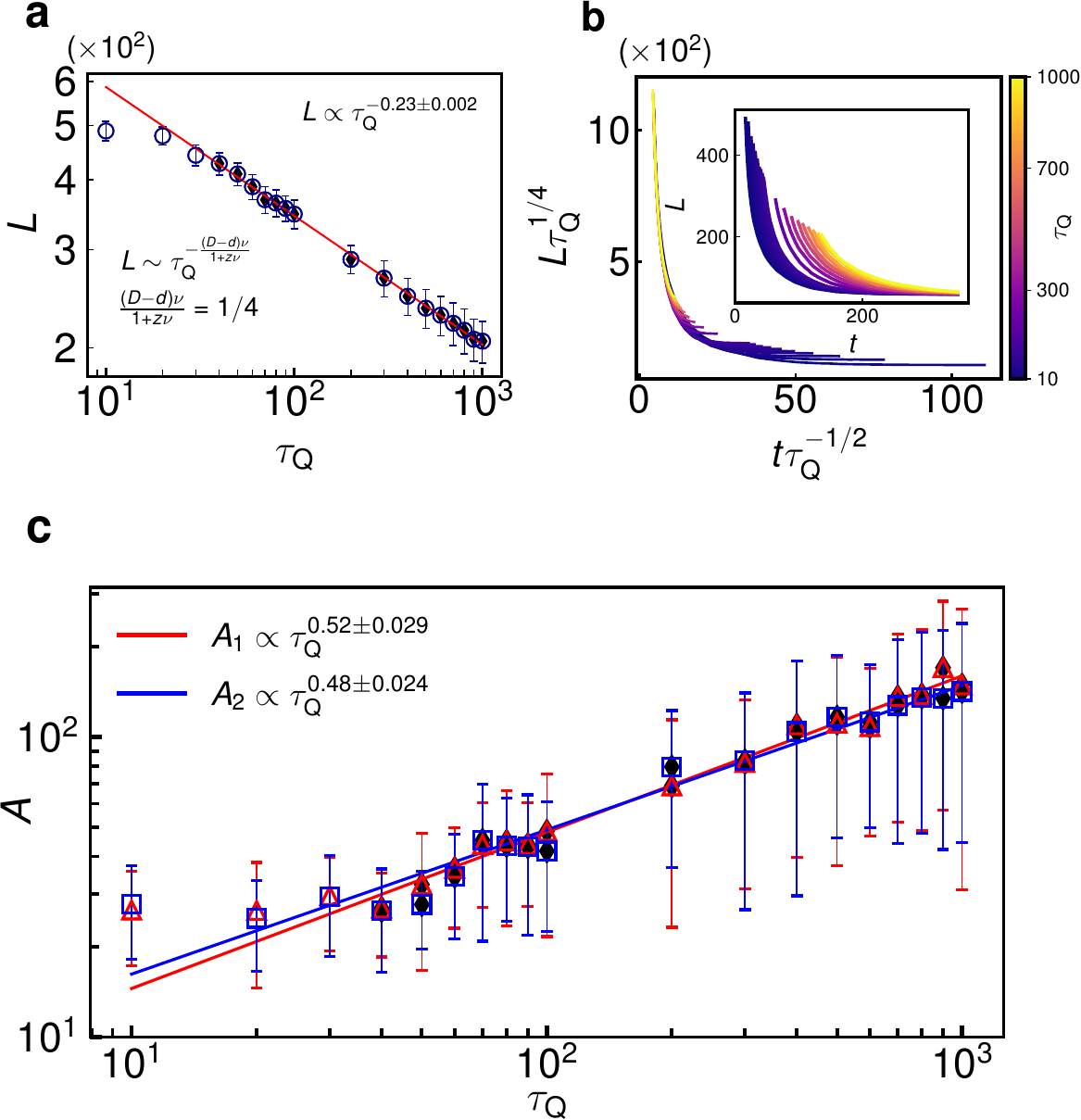}
    \caption{
    \updates{Universality of the domain wall in the immiscible phase.}
    (a) Average domain wall length \(L\) for $|\psi_1|^2$ at 
    \updates{the equilibration time}
    \(
    %t_{\text{eq}}
    t_{\rm eq}
    \) as a function of the quench time \(
    %\tau_Q
    \updates{\tau_{\Q}}
    \) (log scale), averaged over 
    %\(\mathcal{R} = 100\) 
    \updates{100}
    runs, showing 
    \(L \propto 
    %\tau_Q
    \updates{\tau_{\Q}}
    ^{-0.23 \pm 0.002}\). 
(b) 
%Time evolution 
\updates{Evolution}
of \(L\)
%, 
\updates{as a function of time $t$,}
rescaled by \(
%\tau_Q
\updates{\tau_{\Q}}
^{1/4}\), for various \(
%\tau_Q
\updates{\tau_{\Q}}
\); inset shows unscaled 
%\(L(t)\) 
\updates{$L$ as a function of $t$}
from \(
%t_{\text{eq}}
t_{\rm eq}
\). 
(c) Mean domain area \(A\) as a function of \(
%\tau_Q
\updates{\tau_{\Q}}
\), showing 
%KZM 
\updates{Kibble-Zurek}
scaling: 
\(A_1 \propto 
%\tau_Q
\updates{\tau_{\Q}}
^{0.52 \pm 0.029}\), \(A_2 \propto 
%\tau_Q
\updates{\tau_{\Q}}
^{0.48 \pm 0.024}\). 
Error bars show standard deviations over 100 noise realizations.}
    \label{wall_length}
\end{figure}

The system initially develops multiple small domains due to phase separation driven by interspecies repulsion, seeded by non-equilibrium fluctuations. The number of domains \( 
%N_D 
\updates{N_{\rm D}}
\), measured at \( t = t_{\rm eq} \), follows a power-law scaling with the quench time \( 
%\tau_Q 
\updates{\tau_{\Q}}
\), given by \( 
%N_D 
\updates{N_{\rm D}}
\propto 
%\tau_Q
\updates{\tau_{\Q}}
^{-2\nu/(1+z\nu)} \)~\cite{Wu_2017}. The determination of $
%N_D
\updates{N_{\rm D}}
$ from an ensemble of numerical stochastic realizations is discussed 
%in \cite{SM}.
\updates{in the Supplementary Note 2.}
This inverse scaling indicates that slower quenches allow more time near \(t_{\rm eq}\), promoting larger correlated regions and fewer domains. The extracted exponent \( 2\nu/(1+z\nu) \) yields values of 
\( 0.48 \pm 0.017 \) and \( 0.49 \pm 0.024 \)
 for the two species, as shown in Fig.~\ref{density-growth}(d), averaged over 100 noise realizations.  These irregular domains evolve over time $t>t_{\rm eq}$, with the total number $
 %N_D
 \updates{N_{\rm D}}
 $
decreasing via domain coarsening -- a process that merges adjacent regions to reduce interfacial energy by shortening domain 
%walls (shown in the inset of Fig.~\ref{wall_length}(b)). 
\updates{walls.} Each domain wall separates high- and low-density regions, with the total length \( L \) following the KZM scaling 
$L \propto 
%\tau_Q
\updates{\tau_{\Q}}
^{-(D-d)\nu/(1+z\nu)}$
for \( d = 1 \) and \( D = 2 \)~\cite{Ma_2025}. At \( t = t_{\rm eq} \), the fitted exponent $(D-d)\nu/(1+z\nu)$ equals \( 0.23 \pm 0.002 \) and agree well with the theoretical prediction, as shown in Fig.~\ref{wall_length}(a).
Regardless of \( 
%\tau_Q 
\updates{\tau_{\Q}}
\), the system finally relaxes towards a macroscopic phase-separated state with \( L \) saturating to a value dependent on the system size. 
For $t \geqslant 
%t_{\text{eq}}
t_{\rm eq}
$, the domain wall length $L$ scales as $L \sim 
%\tau_Q
\updates{\tau_{\Q}}
^{-\nu/(1+z\nu)} f(t 
%\tau_Q
\updates{\tau_{\Q}}
^{-
z \nu
/(1+z\nu)}) \sim 
%\tau_Q
\updates{\tau_{\Q}}
^{-1/4} f(t 
%\tau_Q
\updates{\tau_{\Q}}
^{-1/2})$, mathematically consistent form with the results in \cite{Ma_2025} and demonstrating universal collapse before the system reaches the equilibrium steady state. This is shown in Fig.~\ref{wall_length}(b).

The mean domain area of the respective components at $t=t_{\rm eq}$ also scales with the quench time $
%\tau_Q
\updates{\tau_{\Q}}
$ as 
$A_1 \propto 
%\tau_Q
\updates{\tau_{\Q}}
^{0.52 \pm 0.029}$ and $A_2 \propto 
%\tau_Q
\updates{\tau_{\Q}}
^{0.48 \pm 0.024}$
 shown in Fig.~\ref{wall_length}(c). The domain characterization, including labeling, mean area, and wall detection, is detailed 
 %in \cite{SM}.
\updates{in the Supplementary Note 2.}

\subsection{Miscible phase and vortex formation}\label{sec3-3}

In the miscible phase (\(g_{12} < g\)), vortices spontaneously emerge as a hallmark of the 
KZM, 
which governs non-equilibrium dynamics after a continuous phase transition. During post-quench evolution, the condensate fragments into protodomains with uncorrelated phases. As domains merge, relative phase differences generate equal number of quantized vortices and antivortices with circulation quantized by the Feynman–Onsager rule~\cite{Onsager1949, Feynman1955, Donnelly1991}. 
\begin{figure*}[!htbp]
    \centering
    \includegraphics[width=0.9\linewidth]{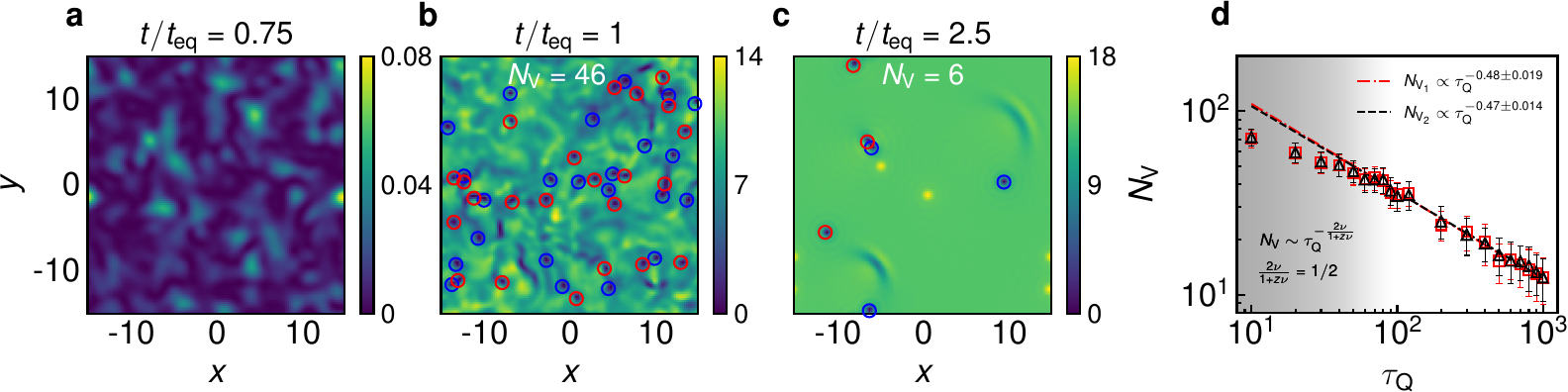}
    \caption{Condensate density evolution %\(|\psi_1|^2\) 
    \updates{of the component 1 of the binary Bose-Bose mixture in the miscible phase.}
    %for \(\tau_Q = 50\).
 Panels (a)–(c) show vortex formation %in the miscible phase 
 \updates{as a function of time $t$ scaled by the equilibration time $t_{\rm eq}$ at the quench time $\tau_{\rm Q} = 50$}, corresponding to red markers in Fig.~\ref{number-growth}(c). 
 Panels (a)–(c) highlight the vortex core formation, with blue/red dots indicating positive/negative winding numbers. 
 \updates{Color bars represent the norm of the condensate order parameter.}
 Panel (d) shows 
 %vortex number scaling: 
 \updates{the scaling of the vortex number $N_{\rm V}:$}
\(
%N_{V_1} 
\updates{N_{\rm V_1}}
\propto 
%\tau_Q
\updates{\tau_{\Q}}
^{-0.48 \pm 0.019}\), \(
%N_{V_2} 
\updates{N_{\rm V_2}}
\propto 
%\tau_Q
\updates{\tau_{\Q}}
^{-0.47 \pm 0.014}\)
, averaged over 100 noise realizations. 
Error bars show standard deviations across these realizations. Grey shaded area in (d) denotes deviation from universal 
%KZM 
\updates{Kibble-Zurek mechanism}
behavior, indicating a smooth crossover between fast and slow quenches.}
    \label{mis-density-growth}
\end{figure*}
For $t \gg t_{\rm eq}$ with any $
%\tau_Q
\updates{\tau_{\Q}}
$ considered here, following vortex–antivortex annihilation, the system relaxes into a low-energy miscible state with complete spatial overlap of the two components.

We determine the vortex number at the equilibration time $
%t_{\text{eq}}
t_{\rm eq}
$, the moment when well-formed vortex cores first become clearly identifiable in the density profile [Fig.~\ref{mis-density-growth}(a)–(c)] marking the maximum in the number of visible vortex cores before subsequently decreasing as a result of annihilation. Notably, in the miscible phase, the vortex or antivortex cores in one component are filled by atoms of the other, forming coreless vortices. The yellow dots in Fig.~\ref{mis-density-growth}(c) signify this and confirm the presence of coreless vortices in $\psi_2$~\cite{wheeler_21,ji_08,gautam_12}. All of these features have been revealed by our simulations. The quench time dependence of the vortex number \( 
%N_V 
\updates{N_{\rm V}}
\) is shown in Fig.~\ref{mis-density-growth}(d), revealing the KZM scaling for point defects (\( d=0 \)) in 2D: 
\( 
%N_{V_1} 
\updates{N_{\rm V_1}}
\propto 
%\tau_Q
\updates{\tau_{\Q}}
^{-0.48 \pm 0.019} \) and \( 
%N_{V_2} 
\updates{N_{\rm V_2}}
\propto 
%\tau_Q
\updates{\tau_{\Q}}
^{-0.47 \pm 0.014} \)
 for the two components.

\subsection{Universal Spatial Statistics}\label{sec3-4}
While KZM predicts universal scaling of the average defect density and domain-wall length with the quench time, it does not specify how defects are arranged in space or whether their spatial correlations exhibit additional universal features. In particular, KZM does not address whether defect positions are correlated, clustered, or randomly distributed at the time of formation. Motivated by recent studies highlighting the role of spatial statistics as an independent probe of nonequilibrium universality~\cite{campo_2022,Thudiyangal_2024,Massaro25migdal}, we therefore analyze the full spatial distributions of vortices in the miscible phase and of mean domain positions in the immiscible phase. By examining distance and nearest-neighbor spacing distributions across many realizations, we test whether defect configurations are consistent with 
PPP 
determined solely by the KZ defect density, thereby extending universality from scaling laws to stochastic-geometric properties of defect patterns.

During growth into the miscible phase, vortices and antivortices emerge at random positions within each superfluid component at \( t \sim t_{\rm eq} \), with positions varying across noise realizations. To reveal emergent spatial correlations, we analyze vortex–vortex and mean domain positions (geometric centroids, 
%see \cite{SM}
\updates{see the Supplementary Note 3}) statistics using a homogeneous 
PPP 
model, where defect density is governed by the KZM 
\cite{campo_2022,Thudiyangal_2024,Massaro25}. 
\begin{figure}[t]
    \centering
    \includegraphics[width=0.7\linewidth]{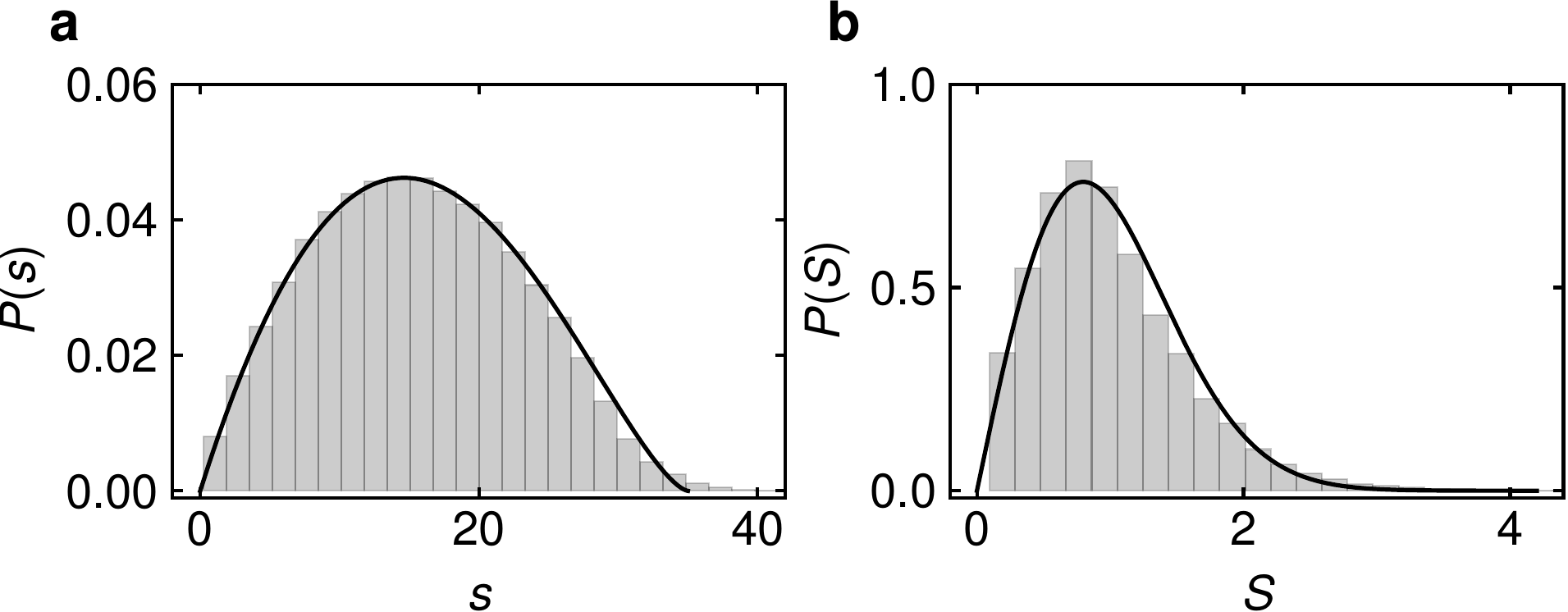}
    \caption{
    \updates{Testing Poisson point process predictions on vortex spatial statistics in the miscible phase.}
    (a) Vortex distance distribution, regardless of the charge, for \updates{the} component 
    %\( j=1 \) 
    \updates{1 of the binary Bose-Bose mixture}
    at 
    \updates{the scaled time}
    \( t/%t_{\text{eq}}
    t_{\rm eq}
    = 0.930 \) with 
    \updates{the quench time}
    \( 
    %\tau_Q 
    \updates{\tau_{\Q}}
    = 50 \)
    %.
    \updates{, where $s$ is the distance and $t_{\rm eq}$ is the equilibration time.}
    The solid line corresponds to the disk line-picking formula with disk radius \( R = 17.5 \) and 25 bins. (b) The first nearest-neighbor spacing distribution of vortices for the same component and quench time, 
    \updates{as a function of the normalized spacing $S = s / \bar{s}$ with $\bar{s}$ being the mean nearest-neighbor distance,}
    fitted with Eq.~\ref{kth-spacing} with $k=1$ (solid line). Histograms are constructed with data from 
    %\( \mathcal{R} = 400 \) 
    \updates{400}
    noise realizations.}
    \label{ppp_dist-vortex}
\end{figure}
Accordingly, the vortex distance distribution \( P(s) \) is expected to follow a uniform distribution over a disk of radius \( R \), described by the disk line-picking formula \cite{Mathai1999,Thudiyangal_2024}: 
\begin{equation}
P(s) = \frac{4s}{
%\pi 
\updates{\rm \pi}
R^2}\left[\arccos\left(\frac{s}{2R}\right) - \frac{s}{2R} \sqrt{1 - \frac{s^2}{4R^2}} \right].
\label{pppeq}
\end{equation}
By aggregating data across all realizations, we smooth out realization-specific randomness and reveal the underlying universal geometric structure of defect positions, which matches well with PPP predictions (Fig.~\ref{ppp_dist-vortex}(a)).

\begin{figure}[t]
    \centering
    \includegraphics[width=0.7\linewidth]{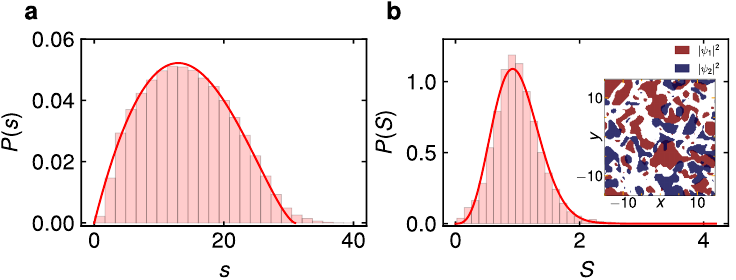}
    \caption{
    \updates{Testing Poisson point process predictions on domain spatial statistics in the immiscible phase.}
    (a) Domain distance distribution for \updates{the} component 
    %\( j=1 \) 
    \updates{1 of the binary Bose-Bose mixture}
    at 
    \updates{the scaled time}
    \( t/
    %t_{\text{eq}}
    t_{\rm eq}
    = 0.930 \) with 
    \updates{the quench time}
    \( 
    %\tau_Q 
    \updates{\tau_{\Q}}
    = 50 \)
    %. 
    \updates{, where $s$ is the distance and $t_{\rm eq}$ is the equilibration time.}
    The solid line corresponds to the disk line-picking formula with disk radius \( R = 15.5 \) and 25 bins.  (b) Further, the first nearest-neighbor spacing distribution for mean domain positions
    \updates{, as a function of the normalized spacing $S = s / \bar{s}$ with $\bar{s}$ being the mean nearest-neighbor distance,}
    follows Eq.~\ref{kth-spacing} with $k=2$ (solid line). The distributions are computed at \( t/
    %t_{\text{eq}}
    t_{\rm eq}
    = 0.930 \) for \( 
    %\tau_Q 
    \updates{\tau_{\Q}}
    = 50 \), using 25-bin histograms constructed from 
    %\( \mathcal{R} = 400 \) 
    \updates{400}
    noise realizations.
(Inset) Density profiles 
\updates{$\left\vert \psi_j \right\vert^2$}
of both components 
\updates{($j = 1, 2$)}
at \( t/
%t_{\text{eq}}
t_{\rm eq}
= 0.930 \), plotted together using a threshold of 50\% of the maximum density (\updates{see the Supplementary Note 2}), from a single noise realization.}
    \label{dist-spacing-ppp-centroid}
\end{figure}

For further characterization, we compute the first nearest-neighbor spacing statistics for vortex positions and mean domain positions in each component~\cite{campo_2022,Thudiyangal_2024}. For each vortex, the probability of finding its nearest neighbor in a shell between \(s\) and \(s+ds\) is evaluated, ensuring that other vortices lie beyond radius \(s\). This is repeated across all vortices and realizations. The $k^{\rm th}$\textemdash order spacing distribution is given by \cite{haake2001,Sakhr_2006}, 
\begin{equation}
    P_k(S) \;=\; \frac{2}{(k-1)!}\, r^{2k}\,
S^{2k-1}\, 
%e
\updates{\e}
^{-r^{2} S^{2}},
\label{kth-spacing}
\end{equation}
where $r= \Gamma(k+\tfrac{1}{2})/\Gamma(k)$. The spacings are normalized by the mean nearest-neighbor distance \(\bar{s}\), and the resulting normalized spacings $S=s/\bar{s}$ give the final distribution.
For vortices, the spacing distribution follows the Wigner-Dyson form $P(S) = \frac{\pi}{2}S \exp\left(-\frac{\pi}{4}S^2\right)$, 
as shown in Fig.~\ref{ppp_dist-vortex}(b) with $k=1$. A similar analysis in the immiscible phase shows that the mean domain positions also follow Poisson statistics (Fig.~\ref{dist-spacing-ppp-centroid}(a)), revealing a comparable geometric structure at the onset of phase separation. Further, the first nearest-neighbor statistics of the mean domain positions yields a spacing distribution, which is in good agreement with the Eq.~\ref{kth-spacing} with $k=2$, shown in the Fig.~\ref{dist-spacing-ppp-centroid}(b). This occurs because the size of domains are finite and also a domain of one component is often separated from its nearest neighbor by a domain of the other component, as seen in the inset of Fig.~\ref{dist-spacing-ppp-centroid}(b). These results reveal strong spatial correlations and universal spacing statistics in both vortex and domain configurations. 
Similar features are displayed by the $j=2$ component, as shown 
%in \cite{SM}. 
\updates{in the Supplementary Note 3.}

Note that two different components in the binary BEC ($\psi_1$ and $\psi_2$) are coupled via the 
inter-species 
interaction coefficient $g_{12}$, affecting each other. Therefore, it is not trivial that the homogeneous PPP can be applied beyond the single-component system, because positions of defects in one component could be affected by positions of defects in other components (as in Eto \textit{et al.} \cite{Eto2011}), possibly breaking ``randomly and independently distributed'' assumption of the homogeneous PPP for positions of defects in each component. Yet, our findings show that our KZM-PPP model is valid even for defects in the binary BEC (component 1 and component 2, respectively), at least around the equilibration time $t_{\rm eq}$ from when defects can be visibly detected in the density profile. 
Indeed, as time goes on, the first nearest neighbor spacing distribution deviates from our KZM-PPP model, as shown 
%in \cite{SM}. 
\updates{in the Supplementary Note 3. }
Still, our results imply that the initial positions of defects in each component could be considered as ``randomly and independently distributed'' points, even if there are interactions between different components.

As a final remark, since our KZM-PPP model does not take into account the detailed form of the interaction between defects, it could also be applied to any multi-component system beyond the BEC.

\section{Conclusions}\label{sec4}

We have shown that symmetry-breaking transitions in a homogeneous binary superfluid, driven by a linear quench, exhibit universal dynamics consistent with the Kibble-Zurek mechanism, which can be extended to account for the fluctuations in the number and spatial distribution of topological defects. Our results can be experimentally tested using atomic mixtures confined in uniform traps, which are ideally suited to probe dynamical scaling laws without causality-induced corrections ubiquitous in inhomogeneous systems \cite{Zurek_2009,delCampo2013,Pyka13,Ulm13,KimShin22}.

We anticipate that the spatial distribution of topological defects in quantum gas mixtures is of relevance to other scenarios, such as the spontaneous emergence of quantum turbulence after a quench \cite{Shinn2025SQT,Massaro25migdal}. In such a context, the incompressible kinetic energy in each component is affected by the vortex positions in the corresponding component, motivating the extension of the previous \cite{Novikov1975,Chavanis2001,Bradley2012QT} results to multi-component systems. A complementary approach concerns the study of the circulation statistics within an arbitrary area in each component and explore extensions of Migdal's area rule \cite{MIGDAL2023,Massaro25migdal}. 
%, e.g., multiferroics, where vortices are created during the ferroelectric transition of hexagonal manganites \cite{Lin14}. 

\section{
%Stochastic formalism
\updates{Methods}
}\label{sec2}

We study the dynamics of symmetry-breaking associated with the emergence of non-zero complex order parameters that pertain to the formation of miscible and immiscible (phase-separated) phases in a steady-state 2D homogeneous Bose-Bose mixture. To establish a reliable theoretical framework that captures the effects of fluctuations in the dynamics of continuous symmetry-breaking phase transitions, we resort to the SPGPE
~\cite{Berlov_2014,Blakie2008,Bradley_2008,Brewczyk_2007,Cockburn2009,bradley_14,Gallucci_2016,Kobayashi_2016,Ota_2018,Proukakis_2006,Proukakis_2008,Proukakis2011,Rooney_2013,Stoof2001,su_2011, RooneySamuelJames2015IaAo, roy_2021,underwood_25},  given by
\begin{eqnarray}
     %i
     \updates{\i}
     \hbar \frac{\partial}{\partial t} \psi_j (\mathbf{x},t) &=& \hat{{\mathcal{P}}} \Bigg\{ (1 - 
     %i
     \updates{\i}
     \gamma) \Bigg[ \Bigg( - \frac{\hbar^2 
     \nabla^2
     }{2 m} + g |\psi_j (\mathbf{x}, t)|^2 
    + g_{12} |\psi_{3 - j}|^2  - \mu(t) \Bigg) \psi_j \Bigg]
    \nonumber\\
    &+& \eta_j (\mathbf{x},t) \Bigg\},\label{eq:sgpe}
\end{eqnarray}
where $\textbf{x} \equiv (x,y)$ and $
\nabla^2
$ are the Cartesian coordinates and Laplacian operator in $2\text{D}$ respectively. 
With $\gamma$ being the dissipation parameter, $\psi_j(\textbf{x},t)$ represents the classical (\textbf{C}) field for each component (species) $j \in {1,2}$, describing the macroscopically occupied low-energy modes subject to random thermal fluctuations represented by the stochastic Gaussian noise term $\eta_j$. The projector $\mathcal{\hat{P}}$ constrains the \textbf{C} field to the coherent region at each time 
%step~\cite{SM}. 
\updates{step (see the Supplementary Methods). }

The intra- and inter-species interaction strengths are denoted by $g$ and $g_{12}$, respectively. Starting from a disordered symmetric phase with $\psi_j = 0$, the system experiences symmetry breaking and transitions to an ordered miscible or immiscible phase depending on the relative values of $g$ and $g_{12}$, as the chemical potential $\mu$ is tuned from below to above the critical point.
We implement a linear ramping of the chemical potential $\mu(t) = 
%\mu_i 
\updates{\mu_{\i}}
+ (t/
%\tau_Q
\updates{\tau_{\Q}}
) (
%\mu_f 
\updates{\mu_{\f}}
- 
%\mu_i
\updates{\mu_{\i}}
)$, 
with $
%\tau_Q
\updates{\tau_{\Q}}
$ being the time of the linear quench. At time $t=0$, the initial chemical potential is $
%\mu_i
\updates{\mu_{\i}}
=0.1$, and is linearly ramped up to the final chemical potential $
%\mu_f
\updates{\mu_{\f}}
=20$ at different quench rates $1/
%\tau_Q
\updates{\tau_{\Q}}
$. We perform numerical simulations to investigate the dynamics of the transition to the miscible $(g_{12}/g=0.5)$ and immiscible $(g_{12}/g=1.05)$ phases with $g = 1$  in a box of size $\mathcal{A} = 30 \times 30 $.
Defining $E_{\rm sc}$ as the energy scale, the length, temperature, and time are measured in units of $\sqrt{\hbar^2/m E_{\rm sc}}$, $E_{\rm sc}/
%k_B
\updates{k_{\rm B}}
$, and $\hbar/E_{\rm sc}$, respectively, where $\hbar$ is the reduced Planck's constant and $
%k_B
\updates{k_{\rm B}}
$ is the Boltzmann constant. 
For convenience, we set $E_{\rm sc} = 
%\mu_{f, p} 
\updates{\mu_{\rm f, p}}
/ 20$ where $
%\mu_{f, p}
\updates{\mu_{\rm f, p}}
$ is the final chemical potential in physical units (around $
%k_B 
\updates{k_{\rm B}}
\times 0.7$nK if one uses the same parameters in the quasi-two-dimensional BEC experiment \cite{Chomaz2015}). 
Additionally, since $\gamma$ does not affect the exponents on the KZ scaling laws \cite{kasamatsu_03,damski_2010}, we set $\gamma = 0.05$.

\backmatter

\bmhead{Supplementary information}
Supplemental material is available for this paper. URL will be inserted later.
% If your article has accompanying supplementary file/s please state so here. 

% Authors reporting data from electrophoretic gels and blots should supply the full unprocessed scans for key as part of their Supplementary information. This may be requested by the editorial team/s if it is missing.

% Please refer to Journal-level guidance for any specific requirements.

\bmhead{Acknowledgements}

We thank Sunil Kumar V. and Sivasankar P.M. for several insightful discussions. Stimulating discussions with Aridaman Singh Chauhan and Subhajit  Paul during the early stages of the work are further acknowledged. 
AR and SP acknowledge the support of the Science and Engineering Research Board (SERB),
Department of Science and Technology, Government of India, under the project
SRG/2022/000057 and IIT Mandi seed-grant funds under the project IITM/SG/AR/87; 
SS and AdC acknowledge financial support from the Luxembourg National Research Fund under Grant No. C22/MS/17132060/BeyondKZM; 
 MT acknowledges the support of Anusandhan National Research Foundation (ANRF), Government of India, through the Prime Minister early career grant ANRF/ECRG/2024/003150/PMS, and Christ University for funding the research through the seed grant, sanction No. CU-ORS-SM-24/94. AR and SP acknowledge the National Supercomputing Mission (NSM) for providing
computing resources of PARAM Himalaya at IIT Mandi, which is implemented by
C-DAC and supported by the Ministry of Electronics and Information Technology
(MeitY) and Department of Science and Technology (DST), Government of India. 

\bmhead{Authors’ Contribution}
MT conceptualized the problem, and SP, MT and AR initiated the project. SP performed all numerical simulations. All authors\updates{, including AdC and SS,} participated in the discussions, data analysis, and writing of the manuscript.

\bmhead{Competing Interests}
The authors declare no competing interests.

\bmhead{Data availability}
The data that support the findings of this study are
available under a "Creative Commons Attribution 4.0 International License" on Zenodo under DOI:\href{https://doi.org/10.5281/zenodo.17608491}{10.5281/zenodo.17608491}. 

\bmhead{Code availability}
Codes are available from the corresponding author upon 
%reasonable request.
% (SHS comment) Editors requested to remove ``reasonable'' in both the Data Availability and Code Availability statement!
\updates{request.}

\bibliography{reference}

\newpage
\pagebreak
\clearpage
%\onecolumngrid
% \begin{center}
% \textbf{\large Supplementary Material for "Kibble-Zurek Scaling and Spatial Statistics in Quenched Binary Bose Superfluids"}
% \end{center}
% \begin{abstract}
%     abcd
% \end{abstract}

%\onecolumngrid
% \section{ABCD}

%\centering
%\section*{Supplementary Material for ``Kibble-Zurek scaling and spatial statistics in quenched binary Bose superfluids''}

\clearpage
\setcounter{figure}{0}
\renewcommand{\thefigure}{\arabic{figure}}
\renewcommand{\figurename}{Supplementary Fig.}

\setcounter{section}{0}
\renewcommand{\thesection}{\arabic{section}}

\renewcommand{\thesubsection}{\arabic{section}.\arabic{subsection}}

\renewcommand{\thesubsubsection}
{S\arabic{section}.\arabic{subsection}.\arabic{subsubsection}}

\begin{center}
\section*{---: Supplementary Material :---}
\end{center}

\begin{center}

{\Large Kibble-Zurek scaling and spatial statistics in quenched binary Bose superfluids}

\vspace{0.6cm}

Subhadeep Patra$^{1}$,
Arko Roy$^{1}$,
Seong-Ho Shin$^{2}$,
Adolfo del Campo$^{2,3}$,
Mithun Thudiyangal$^{4}$

\vspace{0.4cm}

{\small
$^{1}$School of Physical Sciences, Indian Institute of Technology Mandi, Mandi-175005 (H.P.), India\\
$^{2}$Department of Physics and Materials Science, University of Luxembourg, L-1511 Luxembourg, Luxembourg\\
$^{3}$Donostia International Physics Center, E-20018 San Sebasti\'an, Spain\\
$^{4}$Center for Quantum Technologies and Complex Systems (CQTCS), Christ University, Bengaluru, Karnataka 560029, India
}

\vspace{0.3cm}

\end{center}

\section{
%Methodology
\updates{Supplementary Methods}
}
\label{sgpe}
Within the stochastic projected Gross–Pitaevskii equation (SPGPE) formalism, the effective field is constructed by introducing a high-energy cutoff $E_{\rm max}$ and explicitly retaining only the low-energy modes that are relevant for the coherent dynamics of the Bose field~\cite{Gardiner_2003, Blakie2008, Proukakis_2008}. The single-particle modes satisfying $\epsilon_n \le E_{\rm max}$ define the total effective field. Since this space is still too large to be treated fully within a classical-field description~\cite{Blakie2008}, it is further partitioned by introducing an intermediate cutoff energy $\epsilon_{\rm cut}$. This separates the system into a coherent \textbf{C} region, defined by $\textbf{C}={n:\epsilon_n \le \epsilon_{\rm cut}}$, where modes are highly occupied ($\langle n\rangle \sim \mathcal{O}(1)$) and can be treated using the classical-field approximation, and an incoherent \textbf{I} region, defined by $\textbf{I}={n:\epsilon_{\rm cut} < \epsilon_n \le E_{\rm max}}$, which acts as a thermal reservoir giving rise to dissipation and noise terms in the SPGPE. In the \textbf{C} region, quantum fluctuations are neglected in favor of classical field dynamics, while their influence is incorporated indirectly through stochastic coupling to the \textbf{I} region.
The SPGPE formalism thus provides a grand canonical description of the coherent $\mathbf{C}$ field coupled to a thermal reservoir—the incoherent $\mathbf{I}$ region. The coupling strength, controlled by the dissipation parameter $\gamma$, determines the equilibration rate. Thermal fluctuations in the $\mathbf{I}$ region are modeled by Gaussian noise $\eta_j$ satisfying the fluctuation-dissipation relation:
\[
\langle \eta_i(\mathbf{x}, t) \eta_j^*(\mathbf{x}', t') \rangle = 
2 \hbar \gamma 
%k_B 
\updates{k_{\rm B}}
T 
\delta(\mathbf{x} - \mathbf{x}')\delta(t - t')\delta_{ij} 
\]
with $
%k_B
\updates{k_{\rm B}}
$ being the Boltzmann constant and $\langle \cdots \rangle$ denoting noise averaging \cite{roy_2021}.
The $\mathbf{C}$ field $\psi_j(\mathbf{x}, t)$ is restricted to modes below the energy cutoff $\epsilon_{\text{cut}} = 
%k_B 
\updates{k_{\rm B}}
T \ln 2 + \mu$, while modes above belong to the $\mathbf{I}$ region~\cite{Blakie2008, Proukakis_2008, Rooney_2010, Larcher2018, Comaron_2019, Liu_2020}, with $\mu$ and $T$ being its equilibrium chemical potential and temperature. The projector $\hat{\mathcal{P}}$ ensures dynamics remain in the coherent region.
In a homogeneous system, the \textbf{C} field is typically expanded at the initial time as
$\psi(\mathbf{x}, t=0) = \sum_{\mathbf{k}} b_{\mathbf{k}} \, \phi_{\mathbf{k}}(\mathbf{x})$, where
$\phi_{\mathbf{k}}(\mathbf{x}) = \exp({
%i 
\updates{\i}
\mathbf{k}\cdot \mathbf{x}})$ are plane-wave eigenmodes and
$b_{\mathbf{k}}$ are complex amplitudes~\cite{Proukakis_2008}. The cutoff energy $\epsilon_{\rm cut}$
defines the coherent region by restricting the field to low-energy modes, such that modes with
wave vectors satisfying $|\mathbf{k}| \le k_{\rm cut} = \sqrt{2m\epsilon_{\rm cut}}/\hbar$ are retained,
while higher-momentum components are excluded.
In practice, we evolve the field using a pseudospectral method~\cite{Antoine_2013}, starting from
random initial \textbf{C}-field configurations. At each time step, the projection onto the coherent
region is implemented in momentum space by setting Fourier components with
$|\mathbf{k}| \ge k_{\rm cut}$ to zero via fast Fourier transforms. To avoid aliasing errors associated
with the finite spatial grid, however, it is necessary to retain a small number of modes beyond the cutoff, following the standard de-aliasing procedures discussed in \updates{Blakie \textit{et al.}}~\cite{Blakie2008}.

Time evolution of the SPGPE is performed using the forward Euler method, i.e., a finite-difference discretization in time, combined with a spatial discretization on a plane-wave grid. Simulations are carried out in a square box of size 
$L_x \times L_y = 30 \times 30$ with $N_x = N_y = 128$ grid points. The grid spacing satisfies the more stringent than Nyquist criterion for de-aliasing, 
$\Delta x = L_x/N_x < \pi/(2k_{\rm cut})$ (and similarly for $y$)~\cite{Larcher2018}. The temporal step is chosen as $\Delta t \sim 10^{-4}$, ensuring numerical stability and convergence. The equilibrium atom number obtained in the simulations is consistent with the target final chemical potential $
%\mu_f
\updates{\mu_{\rm f}}
$. We estimate that the total computational cost of the simulations presented here exceeds $1.5 \times 10^{4}$ CPU hours.

We study transitions to miscible ($g_{12}/g = 0.5$) and immiscible ($g_{12}/g = 1.05$) phases with $g = 1$, $T = 10^{-6}$, $\gamma = 0.05$, and $\epsilon_{\text{cut}} = 
%k_B 
\updates{k_{\rm B}}
T \ln 2 + 
%\mu_f
\updates{\mu_{\f}}
$.

\section*{\updates{Supplementary Note 1: }
Finding the equilibration time $t_{\rm eq}$}  
\label{teq}
In our work, we identify scaling laws for various observables evaluated at the 
equilibration time $
%t_{\text{eq}}
t_{\rm eq}
$, extracted for a range of finite quench time $
%\tau_Q
\updates{\tau_{\Q}}
$. Determining $
%t_{\text{eq}}
t_{\rm eq}
$ is nontrivial, as it marks the crossover from the exponential growth to the adiabatic linear growth of the scaled norm $\mathcal{N}$ of the order parameter coinciding with the onset of vortex proliferation and domain merging.

\begin{figure}[!htbp]
    \centering
    \includegraphics[width=0.7\linewidth]{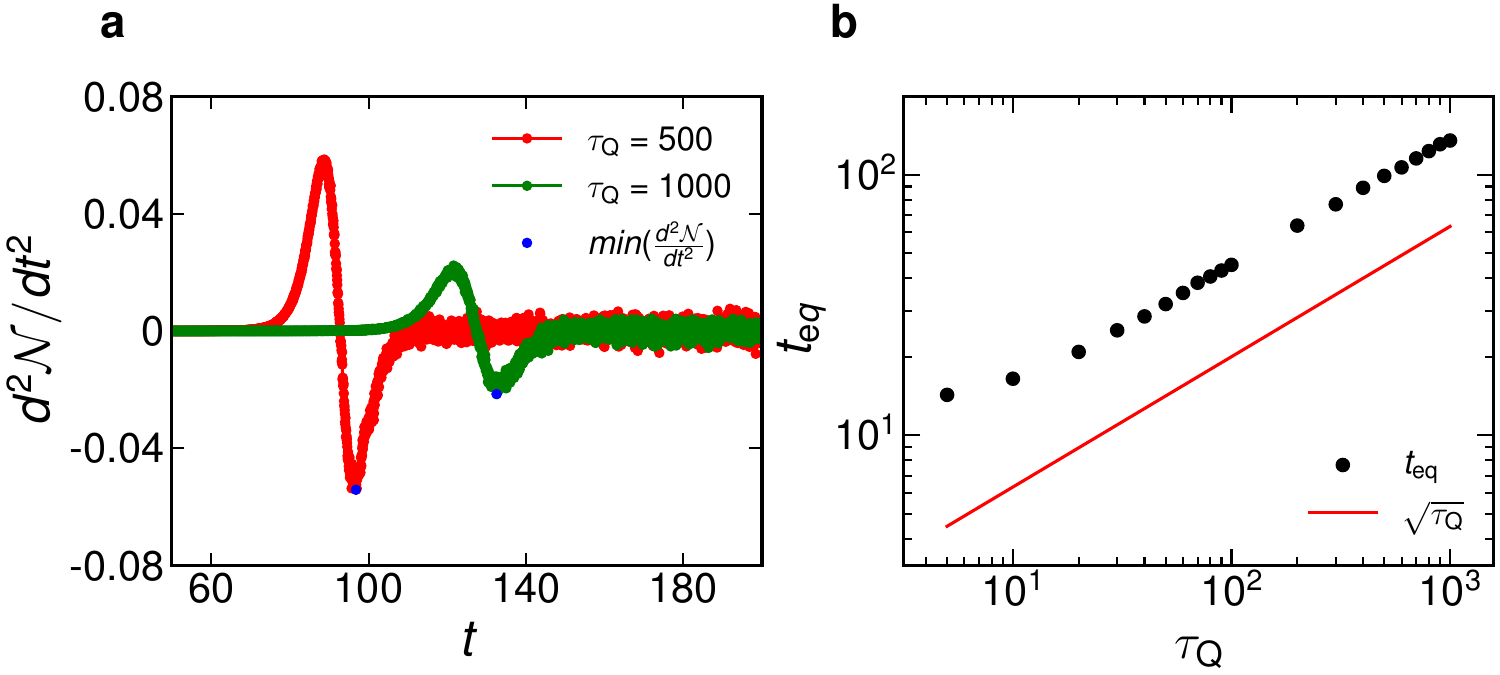}
    \caption{
    Determination of equilibration time %$t_{\text{eq}}$ 
    for various quench \updates{times} %time $\tau_Q$ 
    in the immiscible phase\updates{, from the time evolution of the condensate norm.}
    (a) Second derivative of the condensate norm $d^2 \mathcal{N}/dt^2$ shown for two representative quench %time, 
    \updates{times,}
    $
    %\tau_Q
    \updates{\tau_{\Q}}
    = 500$ and $1000$. Blue markers indicate the minima of $d^2 \mathcal{N}/dt^2$, used as reference points for estimating 
    \updates{the equilibration time}
    $
    %t_{\text{eq}}
    t_{\rm eq}
    $. (b) Extracted 
    equilibration times $
    %t_{\text{eq}}
    t_{\rm eq}
    $ as a function of $
    %\tau_Q
    \updates{\tau_{\Q}}
    $. The theoretical scaling $
    %t_{\text{eq}} 
    t_{\rm eq}
    \propto \sqrt{
    %\tau_Q
    \updates{\tau_{\Q}}
    }$ is shown for comparison. Deviations at faster quenches suggest a breakdown of Kibble-Zurek scaling in the high-speed regime.}
    \label{find_t_eq}
\end{figure}

To locate this crossover point, we first compute the second derivative 
$d^2 \mathcal{N}/dt^2$ across time for each $
%\tau_Q 
\updates{\tau_{\Q}}
\in [5, \cdots, 1000]$, as shown in Fig.~\ref{find_t_eq}(a). The minimum of this second derivative provides a characteristic time, but this does not precisely coincide with the true crossover point where protodomains begin to merge. To bridge this gap, we apply a constant time shift to these minima to estimate $
%t_{\text{eq}}
t_{\rm eq}
$.

For quench time in the range $
%\tau_Q
\updates{\tau_{\Q}}
= 5$ to $100$, the amount of shift is fixed and determined using the dynamics from the fastest quench. For slower quenches ($
%\tau_Q
\updates{\tau_{\Q}}
= 200$ to $1000$), we apply a larger time shift. This distinction accounts for the increased time required for domain interactions at slower quenches, as observed in the density evolution profiles.

We note that the numerical computation of second derivatives is sensitive to the temporal resolution and smoothness of the data, and the manual shift is limited by the discretization of the simulation time steps. Nevertheless, this procedure is consistently applied to both miscible and immiscible regimes. Figure~\ref{find_t_eq}(b) shows the extracted values of $
%t_{\text{eq}}
t_{\rm eq}
$ as a function of $
%\tau_Q
\updates{\tau_{\Q}}
$. These follow the expected scaling behavior $
%t_{\text{eq}} 
t_{\rm eq}
\propto \sqrt{
%\tau_Q
\updates{\tau_{\Q}}
}$, in agreement with Kibble-Zurek (KZ) theoretical predictions.

\begin{figure}[t]
    \centering
    \includegraphics[width=0.7\linewidth]{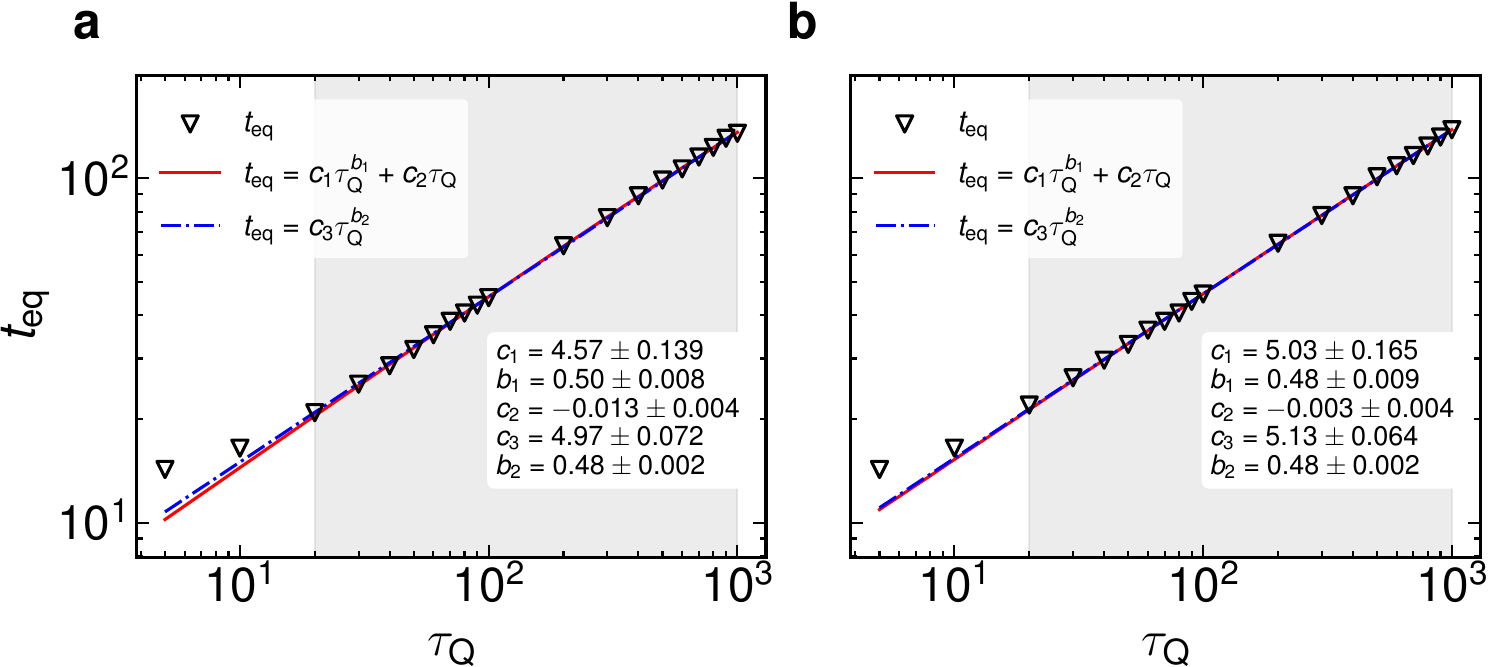}
    \caption{
    \updates{Testing the Kibble-Zurek prediction on the equilibration time.}
    Scaling of the equilibration time $
    %t_{\text{eq}}
    t_{\rm eq}
    $ with the quench time $
    %\tau_Q
    \updates{\tau_{\Q}}
    $ for (a) the immiscible phase and (b) the miscible phase. In panel (a), the data are fitted both with a time-shift correction, $
    %t_{\text{eq}}
    t_{\rm eq}
    = c_1 
    %\tau_Q
    \updates{\tau_{\Q}}
    ^{b_1} + c_2 
    %\tau_Q
    \updates{\tau_{\Q}}
    $, and without the shift, $
    %t_{\text{eq}}
    t_{\rm eq}
    = c_3 
    %\tau_Q
    \updates{\tau_{\Q}}
    ^{b_2}$. Panel (b) shows the corresponding scaling behavior for the miscible phase.}
    \label{fitting_teq}
\end{figure}

%In the SPGPE framework, the critical chemical potential satisfies $\mu_c \ne 0$, and it is difficult to calculate $\mu_c$ at nonzero temperature, making it non-trivial $\mu = \mu_c$ at $t = 0$ \cite{Liu_2020}.
In the standard Kibble-Zurek mechanism (KZM) 
picture, the freeze-out time is defined relative to the instant the system crosses 
\updates{the critical chemical potential}
$
%\mu_c
\updates{\mu_{\rm c}}
$ at $t=0$. 
\updates{
However, in the SPGPE framework, the symmetry-breaking transition happens at some positive chemical potential $\mu_{\rm c}$, whose value is difficult to calculate \cite{Liu_2020}. Therefore, we cannot make the system cross the critical point at $t = 0$. 
Instead, we introduce the time shift $t_{\rm c}$, satisfying $\mu \left( t_{\rm c} \right) = \mu_{\rm c}$, as a fitting parameter to test KZ scaling. This approach is valid as shown in \cite{Liu_2020} and is easier to test KZ scaling than making the transition occur at $t = 0$ by exactly calculating $\mu_{\rm c}$.}
To account for this shift, we evaluate the equilibration time $
%t_{\text{eq}}
t_{\rm eq}
$ relative to $
%t_c
\updates{t_{\rm c}}
$, such that the effective time $(
%t_{\text{eq}} 
t_{\rm eq}
- 
%t_c
\updates{t_{\rm c}}
)$ scales with the quench time $
%\tau_Q
\updates{\tau_{\Q}}
$ as
\begin{equation}
%t_{\text{eq}} 
t_{\rm eq}
= c_1 
%\tau_Q
\updates{\tau_{\Q}}
^{b_1} + c_2 
%\tau_Q
\updates{\tau_{\Q}}
,
\end{equation}
where $c_2 = (
%\mu_c
\updates{\mu_{\rm c}}
- 
%\mu_i
\updates{\mu_{\i}}
)/(
%\mu_f
\updates{\mu_{\f}}
- 
%\mu_i
\updates{\mu_{\i}}
)$. For the immiscible phase, the fitting yields $c_1 = 4.57 \pm 0.139$, $b_1 = 0.50 \pm 0.008$, and $c_2 = -0.013 \pm 0.004$, as shown in Fig.
\ref{fitting_teq}(a). 
For comparison, we also fit the data without including the shift, using
\begin{equation}
%t_{\text{eq}}
t_{\rm eq}
= c_3 
%\tau_Q
\updates{\tau_{\Q}}
^{b_2},
\end{equation}
which gives $c_3 = 4.97 \pm 0.072$ and $b_2 = 0.48 \pm 0.002$. The exponent $b_2$ is in excellent agreement with the KZM prediction, as evident from Fig.~\ref{fitting_teq}(a). A similar analysis performed in the miscible phase leads to the fits shown in Fig.~\ref{fitting_teq}(b).

\section{\updates{Supplementary Note 2: }
Characterization of domains}
\label{domains}
\subsection{Domain labeling}

To analyze domain formation and evolution in the immiscible regime ($g_{12} > g$) for times $t \geqslant t_{\rm eq}$, we convert the numerically obtained density profiles into binary images using a threshold-based method. Specifically, at each time step, we apply a threshold equal to $50\%$ of the maximum density in each component: $
%n_{b_1} 
\updates{n_{\rm b_1}}
= 0.5 \times n_{1, \text{max}}$ and $
%n_{b_2} 
\updates{n_{\rm b_2}}
= 0.5 \times n_{2, \text{max}}$. Any pixel ({\tt px}) with a local density greater than the threshold is assigned a binary value of 1, and 0 otherwise. Mathematically, this is expressed as:
\begin{equation}
    P = \begin{cases}
    1, & 
    %n_1 > n_{b_1}
    \updates{n_j > n_{{\rm b}_j}}
    , \\
    0, & 
    %n_1 < n_{b_1}
    \updates{n_j < n_{{\rm b}_j}}
    ,
    \end{cases}
    \label{binc}
\end{equation}
\updates{for $j = 1, 2$. }

\begin{figure*}[!htbp]
    \centering
    \includegraphics[width=0.85\linewidth]{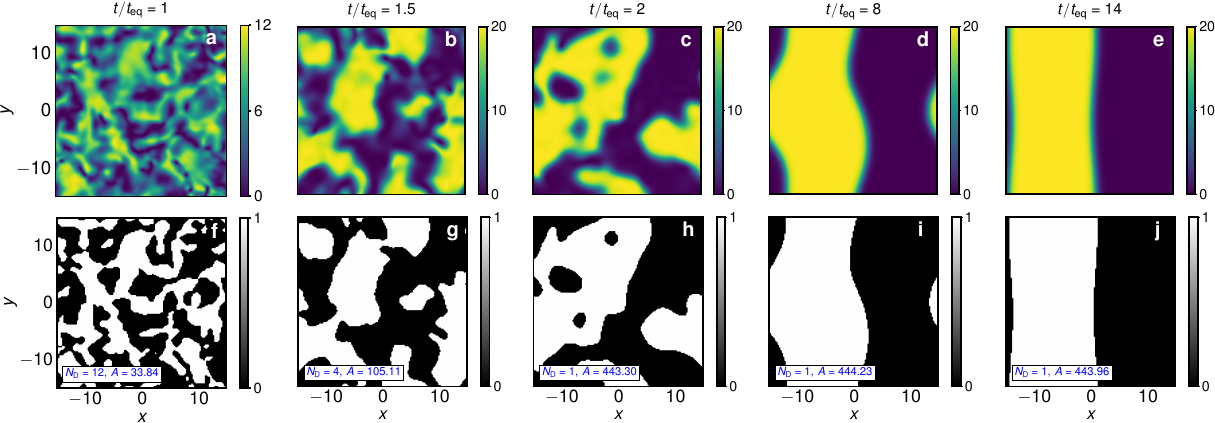}
    \caption{Time evolution of the density %\( n_1 = |\psi_1|^2 \) 
    \updates{of the component 1}
    of 
    %a 
    \updates{the}
    binary Bose-Bose mixture in the immiscible phase.
    %(\( g_{12} > g \)) for the quench time \( \tau_Q = 50 \)
    \updates{Here, the time $t$ is scaled in units of the equilibration time $t_{\rm eq}$, at the quench time $\tau_{\Q} = 50$.}
    (a)--(e) Density profiles of one component of the binary mixture are shown at successive time steps starting from the time \( 
    %t_{\text{eq}}
    t_{\rm eq}
    \), illustrating the evolution of domain structures. 
    \updates{Color bars represent the norm of the condensate order parameter.}
    (f)--(j) Corresponding binary images obtained from the density profiles in (a)--(e), using a fixed threshold criterion Eq.~\ref{binc} and plotted in grayscale. These binary fields are used to identify and analyze domain formation. 
    \updates{$N_{\rm D}$ represents the total number of domains.}
}
    \label{binary_density}
\end{figure*}

This binary representation distinguishes between domain regions ($P = 1$) and background or complementary regions ($P = 0$). From the original density profiles shown in Figs.~\ref{binary_density}(a)--(e),  the resulting binary images for component $n_1$ are displayed in Figs.~\ref{binary_density}(f)--(j). The threshold is chosen to ensure that the number of domains and mean domain area (discussed later) at the final equilibrium state are consistent with physical expectations. That is, in the steady state, density  $n_j$ occupies the simulation domain area ${\mathcal A}/2$, as shown in Figs.~\ref{binary_density}(e, j). 
To identify and count the domains, we employ an image-processing method based on the \textit{8-connected component labeling scheme}~\cite{haralick_1992,gonzalez_2018}. This approach examines each pixel with $P = 1$ and considers its eight neighboring pixels (top, bottom, left, right, and diagonals). All pixels with $P=1$ that are connected through their surrounding positions are grouped into clusters and assigned unique labels, thus identifying distinct domains. A schematic demonstration of this procedure is shown in Fig.~\ref{domain-labeling}(a-c), where a representative density field using a threshold value 
\updates{is applied}
to illustrate the labeling mechanism.
In simulations with periodic boundary conditions, domains that wrap around the edges are merged to ensure continuity. After completion of the labeling, the total number of domains $
%N_D
\updates{N_{\rm D}}
$ is obtained by counting the distinct regions labeled in the binary field. We eventually take an ensemble average of $
%N_D
\updates{N_{\rm D}}
$ over different noise realizations to obtain the KZ scaling laws.

\subsection{Mean area of domains}
\label{mad}
\begin{figure}[!htbp]
    \centering
    \includegraphics[width=0.7\linewidth]{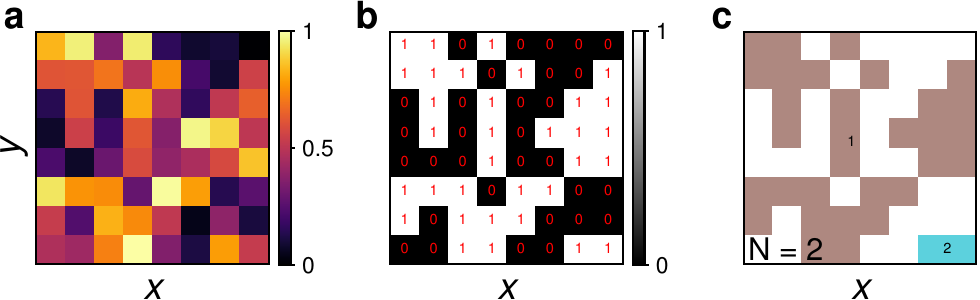}
    \caption{\updates{Schematic illustration of the domain detection method.}
    (a) A sample density field where pixel intensity represents the local density. (b) Binary conversion of the density field using a threshold value of 0.5. Pixels with density greater than the threshold are assigned a value of 1 (white), while the rest are set to 0 (black), representing the background. (c) Identification of domains using the 8-connected component labeling scheme, where each connected region with value 1 is assigned a unique label and visualized in distinct colors. The total number of such labeled regions is denoted by \( N \), representing the number of detected domains. Note that periodic boundary conditions are not applied in this schematic for clarity.
}
    \label{domain-labeling}
\end{figure}

To quantify the size of each domain, we calculate its area by counting the number of pixels associated with a given labeled object, as shown in Fig.~\ref{domain-labeling}. Let \( N_i \) denote the number of pixels that belong to the \( i^\text{th} \) domain. Each pixel corresponds to an area \( 
%A_p 
\updates{A_{\rm p}}
= \mathcal{A} / N_{\text{grid}} \), where \( \mathcal{A} \) is the total area of the simulation box and \( N_{\text{grid}} \) is the total number of points in the grid. The area of the \( i^\text{th} \) domain is then given by $A_i = N_i \times 
%A_p
\updates{A_{\rm p}}
.$
The mean domain area is computed as
\begin{equation}
    A = \frac{1}{
    %N_D
    \updates{N_{\rm D}}
    } \sum_{i=1}^{
    %N_D
    \updates{N_{\rm D}}
    } A_i,
\end{equation}
where $
%N_D
\updates{N_{\rm D}}
$ is the total number of domains detected for one of the components. After ensemble averaging, this mean domain area $A$ provides a measure of the typical domain size in the system. As discussed in the main text, tracking the time evolution of $A$ plays a key role in extracting the KZ 
scaling laws during the non-equilibrium dynamics of the system. 
\begin{figure}[!htbp]
    \centering
    \includegraphics[width=0.6\linewidth]{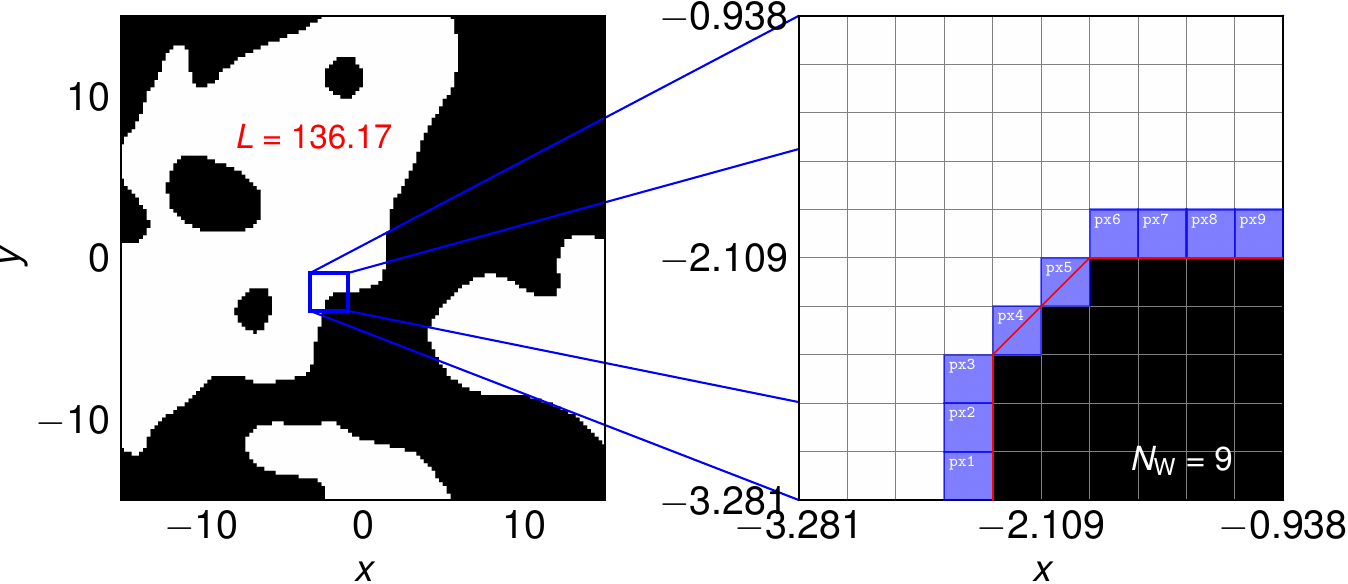}
    \caption{
    \updates{Schematic illustration of the domain wall detection method.}
    A zoomed-in view of a small region of the system is used to illustrate how domain walls are identified at the pixel level. In this representation, black pixels correspond to \( P = 0 \) (background), and white pixels correspond to \( P = 1 \) (domain). Pixels that lie at the interface---where a transition occurs between \( P = 1 \) and \( P = 0 \) in their immediate neighborhood---are marked in blue. These are identified as boundary pixels contributing to the domain wall. In this example, there are nine such boundary pixels, labeled \texttt{px1} to \texttt{px9}, where the change from domain to background is detected. The panel shown is based on a segment from Fig.~\ref{binary_density}(h), used here for visualizing the domain wall detection scheme.
}
    \label{wall_pixel}
\end{figure}

\subsection{Domain wall detection}
\label{dwd}
Once the binary density field is obtained, the domain walls are identified as the boundaries that separate regions with different binary values. Specifically, a domain wall pixel is one that has a value \( P = 1 \) and is adjacent to at least one neighboring pixel with \( P = 0 \). This change in the binary value across neighboring pixels signals the presence of a domain boundary. We systematically scan the binary grid and mark such pixels as part of the domain wall, as illustrated in 
%Fig.~\ref{wall_pixel}(a).
\updates{
Fig.~\ref{wall_pixel}. 
}

To quantify the total domain wall length, we count the number of these boundary pixels and convert them into physical units using the spatial resolution \( \Delta x \), which is the step size of the grid. For pixels that are connected to neighboring boundary pixels along the horizontal or vertical directions, the contribution to wall length is simply \( \Delta x \). However, for pixels that are connected diagonally on both sides, the contribution is \( \sqrt{2} \Delta x \), as shown in 
%Fig.~\ref{wall_pixel}(b), 
\updates{
Fig.~\ref{wall_pixel}, 
}
specifically for the pixels labeled \( \texttt{px4} \) and \( \texttt{px5} \). The total domain wall length is then given by:
\begin{equation}
    L = N_{\text{side}} \times \Delta x + N_{\text{diagonal}} \times \sqrt{2} \Delta x,
\end{equation}
where \( N_{\text{side}} \) is the number of boundary pixels connected via side (horizontal or vertical) neighbors, and \( N_{\text{diagonal}} \) is the number connected via diagonal neighbors. The total number of boundary pixels is \( N_{\text{w}} = N_{\text{side}} + N_{\text{diagonal}} \). The effective domain wall length corresponds to the red line traced along the edge pixels in 
%Fig.~\ref{wall_pixel}(b).
\updates{
Fig.~\ref{wall_pixel}. 
}

This method ensures that only those pixels at the interface between \( P = 1 \) and \( P = 0 \) are identified, and the physical length of the domain wall is accurately computed by considering the geometry of pixel connectivity.

\section{\updates{Supplementary Note 3: }
Universal spatial statistics and geometric centroids}
\label{dstat}
\begin{figure}[!htbp]
    \centering
    \includegraphics[width=0.65\linewidth]{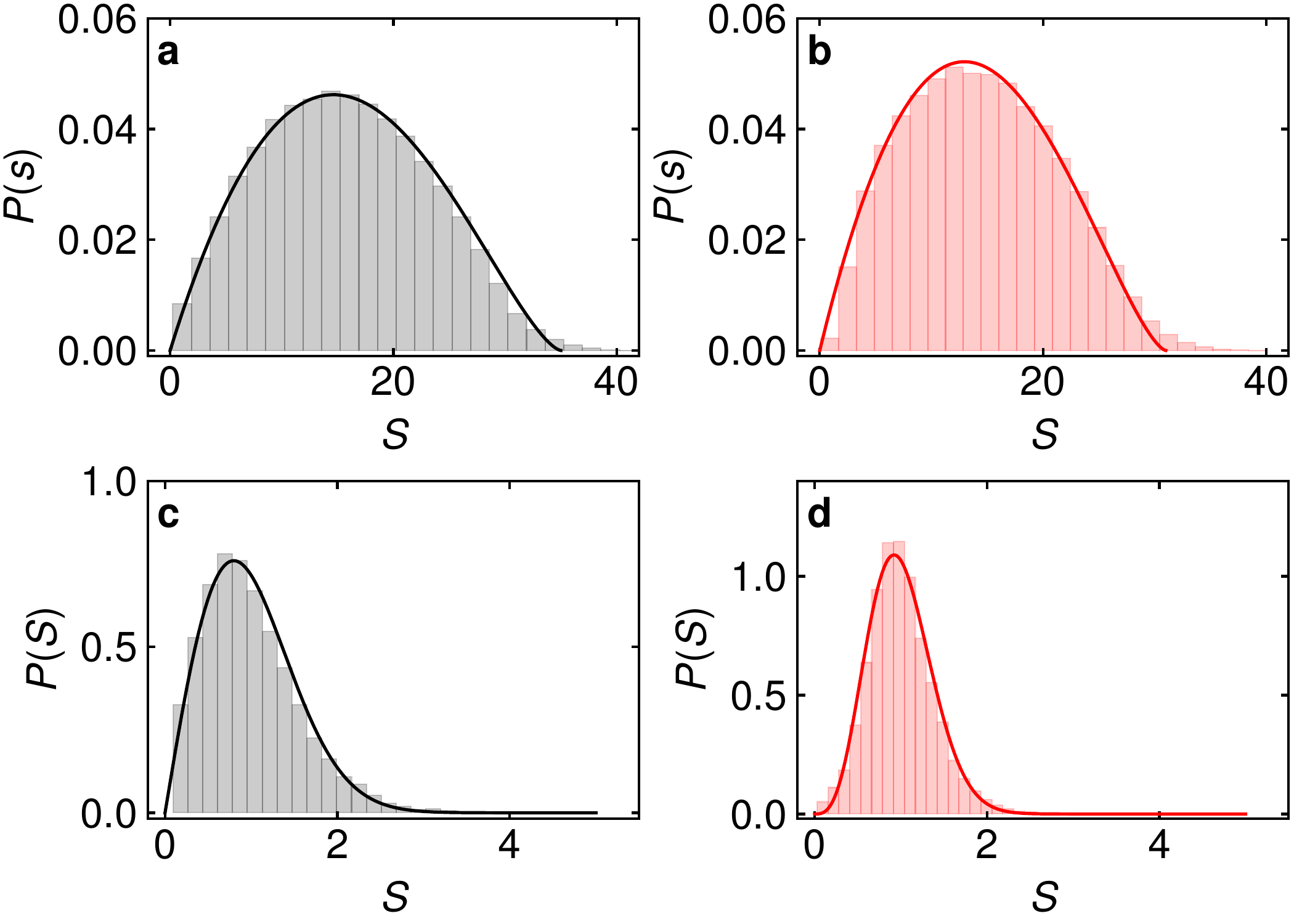}
    \caption{
    \updates{Testing Poisson point process predictions on vortex and domain spatial statistics for the component 2 of the binary Bose-Bose mixture.}
    (a) Vortex distance distribution %(component $j=2$) 
    at 
    \updates{the scaled time}
    $t/t_{\rm eq} = 0.930$ for 
    \updates{the quench time}
    $
    %\tau_Q
    \updates{\tau_{\Q}}
    =50$, fitted with Eq.~
    %6
    \updates{5}
    (main text))
    %. 
    \updates{where $s$ is the distance and $t_{\rm eq}$ is the equilibration time.}
    (b) Distance distribution of mean domain positions (geometric centroids), also fitted with Eq.~
    %6 
    \updates{5}
    (main text). (c) First nearest-neighbor vortex spacing distribution, fitted with the Wigner–Dyson formula
    %. 
    \updates{where $S = s / \bar{s}$ is the distance normalized by the mean nearest-neighbor distance $\bar{s}$.}
    (d) Spacing for mean domain positions %(component $j=2$) 
    fitted with the Eq.~
    %7
    \updates{6}
    (main text) with $k=2$. All distributions are computed at $t/t_{\rm eq} = 0.930$ for $\tau_Q=50$ using 25-bin histograms averaged over 
    %$\mathcal{R}=400$ 
    \updates{400}
    trajectories.}
    \label{ppp-wigner}
\end{figure}
\begin{figure}[!htbp]
    \centering
    \includegraphics[width=0.6\linewidth]{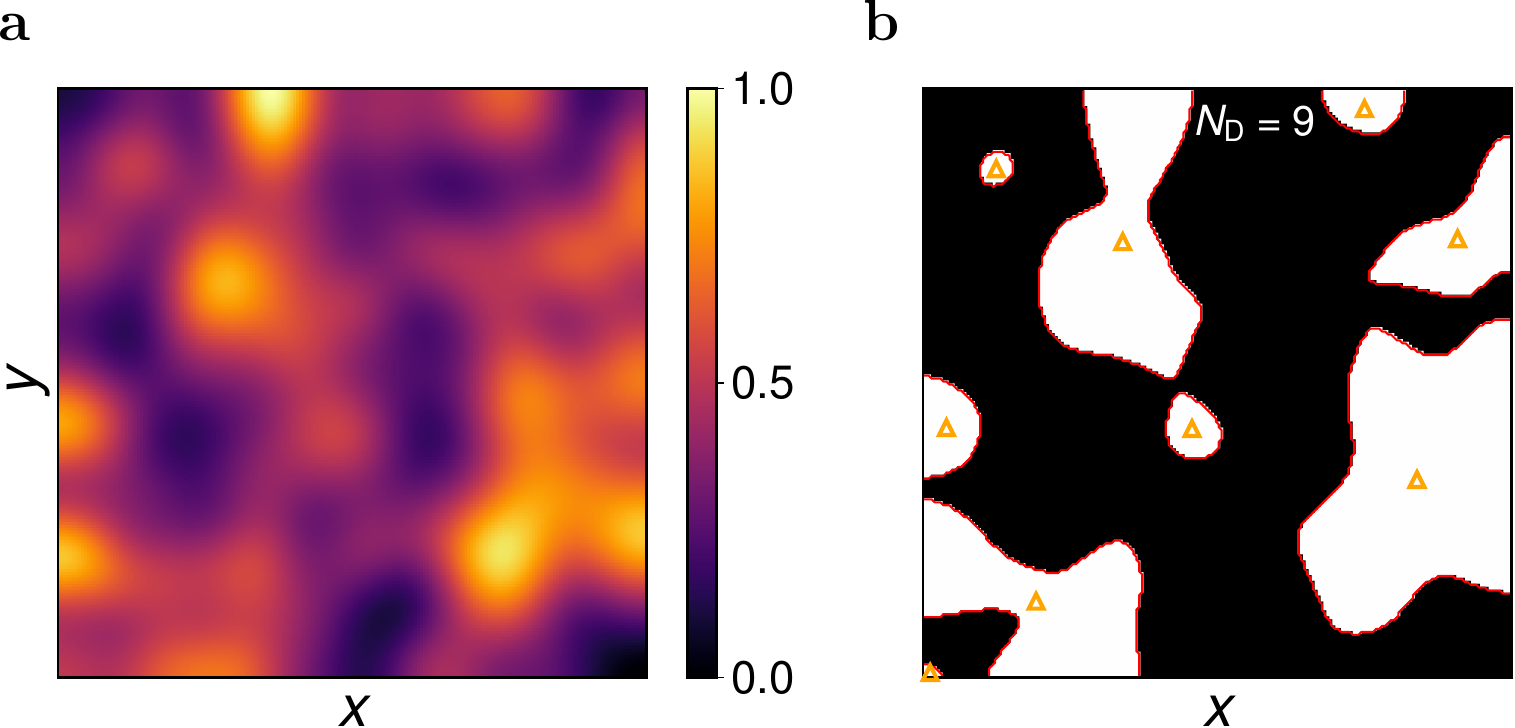}
    \caption{Schematic illustration of domains and its geometric centroids (mean domain positions) without considering the periodic boundary conditions.
    Panel (a) represents a density profile with density variation from $0$ to $1$ and panel (b) represents the binary field of the density considering a threshold value $0.5$ i.e $50\%$ of the maximum density. The orange markers represent the locations of the centroids of the binary domains.}
    \label{centroids_schematic}
\end{figure}
During condensate growth, defects form at the interfaces of merging protodomains with a finite probability. The number of such candidate locations scales as $
%\mathcal{N}_D
\updates{\mathcal{N}_{\rm D}}
\sim \mathcal{A}/\hat{\xi}^2$, where $\mathcal{A}$ is the system size and $\hat{\xi}$ is the out-of-equilibrium correlation length. Assuming independent defect formation, $
%\mathcal{N}_D
\updates{\mathcal{N}_{\rm D}}
$ sites follow a binomial distribution \cite{Ruiz_2020}, and the average number of defects scales as
\begin{equation}
\langle N \rangle \sim 
%\tau_Q
\updates{\tau_{\Q}}
^{-2\nu/(1+z\nu)},
\end{equation}
consistent with the KZ scaling shown in 
%Fig.~2(d) (main text) and Fig.~2(h) (main text). 
\updates{Fig.~3(d) (main text) and Fig.~5(d) (main text).}
As mentioned in the main text, the spatial arrangement of defects can be modeled as a homogeneous Poisson point process (PPP). To verify this even for $j=2$, we study the pairwise distance distribution $P(s)$ between defects, also known as the disk-line picking distribution in 2D. The numerical $P(s)$ results for vortices [Fig.~\ref{ppp-wigner}(a)] and mean domain positions (treated here as effective point defects, $d=0$)[Fig.~\ref{ppp-wigner}(b)] agree well with the theoretical prediction (Eq.~
%6
\updates{5 }
(main text)).
Further insight is obtained from nearest-neighbor statistics. The $k$th-order spacing distribution for a 2D PPP is given by Eq.~
%7 
\updates{6 }
(main text).
Note that for $k=1$, this reduces to the nearest-neighbor distribution $P(S)$, which coincides with the Wigner surmise for level repulsion in random matrix theory with $\beta=1$~\cite{Guhr1998, Sakhr_2006}. Our calculations of first-nearest neighbor spacing for vortices confirm this: the numerical distributions are in good agreement with Eq.~
%7 
\updates{6 }
(main text) for $k=1$, as shown in Fig.~\ref{ppp-wigner}(c). 
\begin{figure*}[!h]
    \centering
    \includegraphics[width=0.85\linewidth]{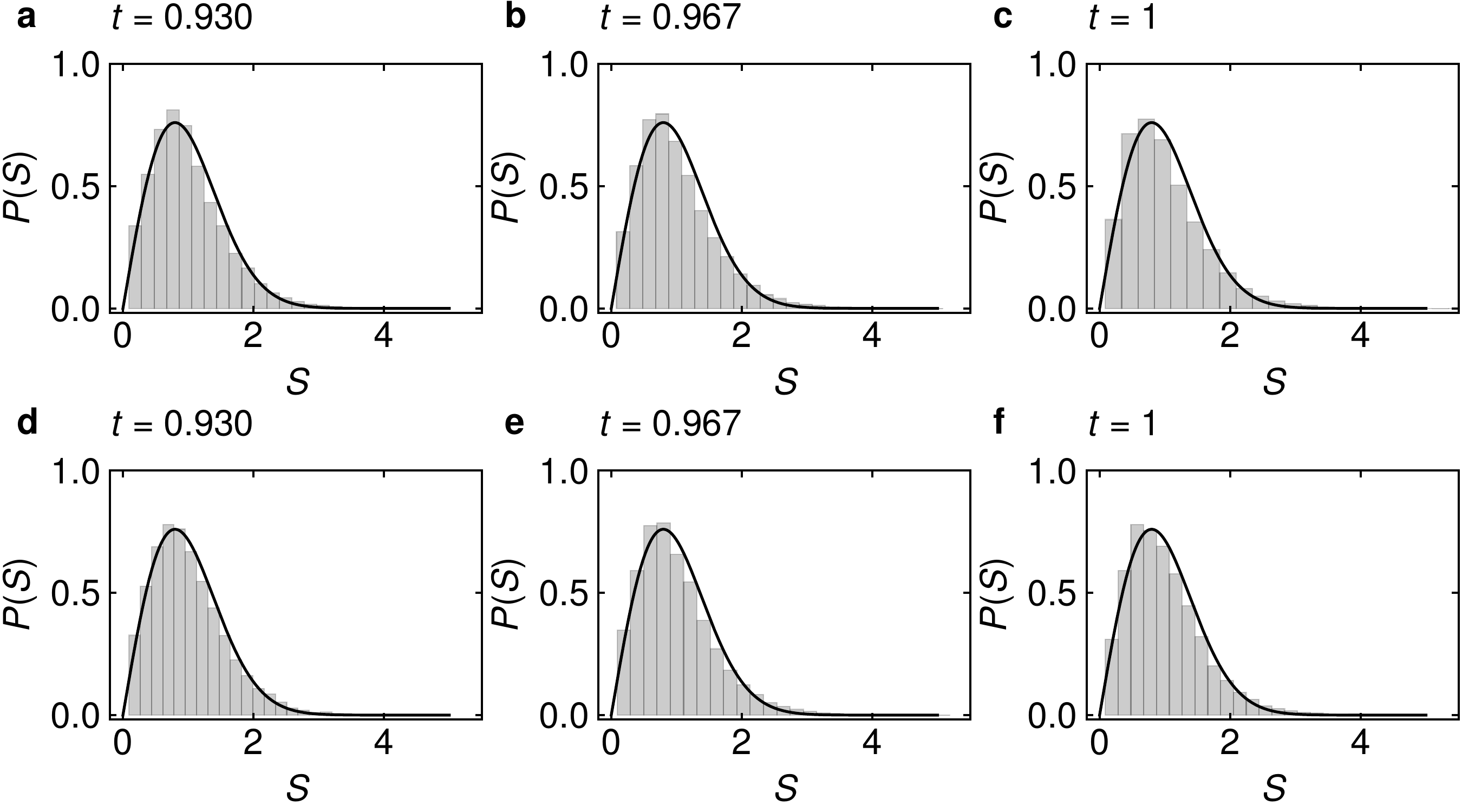}
    \caption{\updates{Time evolution of the first} 
    %First
    -nearest-neighbor spacing distributions of vortices in the miscible phase.
    %for quench time \( \tau_Q = 50 \), shown at \( t/t_{\rm eq} = 0.930 \), \( 0.967 \), and \( 1.0 \). 
    \updates{The quench time $\tau_{\Q}$ is fixed to $\tau_{\Q} = 50$. Here, $S$ is the distance scaled by the mean nearest-neighbor distance, and $t / t_{\rm eq}$ is the scaled time in units of the equilibration time $t_{\rm eq}$.}
    \updates{Panels (a--c) correspond to component
    %\( j=1 \) and \( j=2 \) 
    1 of the binary Bose-Bose mixture}. Solid black lines denote the theoretical distribution from Eq.~
    %7 
    \updates{6}
    (main text) with \( k=1 \). \updates{Panels (d--f) correspond to component 2 of the mixture, with the black line indicating the theoretical distribution given in  the Eq.~ 6 (main text), with \( k =1 \).}  Histograms are constructed from 
    %$\mathcal{R}=400$ 
    \updates{400}
    noise realizations with 25 bins.}
    \label{vortex_spacing}
\end{figure*}
\begin{figure*}[!h]
    \centering
    \includegraphics[width=0.85\linewidth]{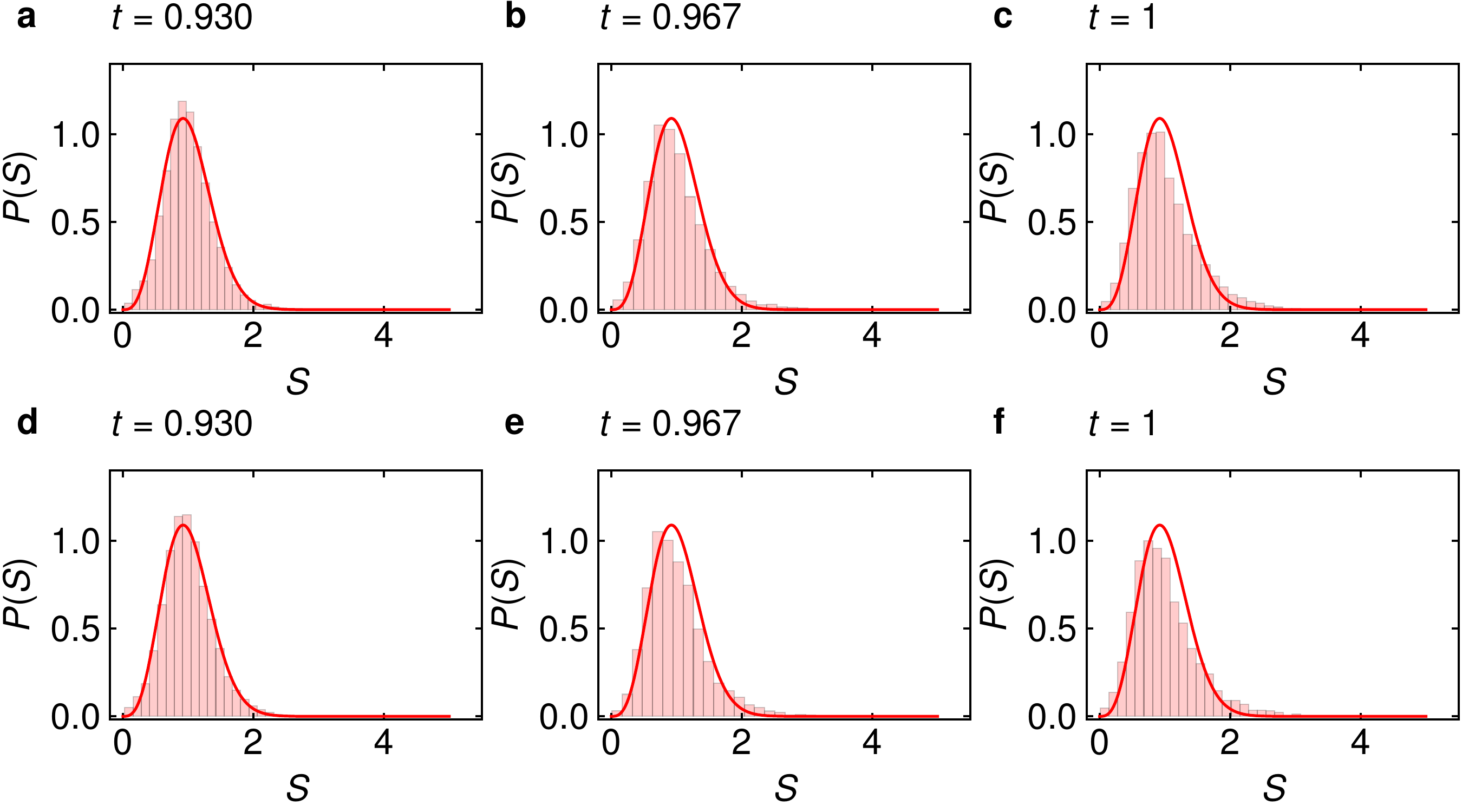}
    \caption{\updates{Time evolution of the first}
    %First
    -nearest-neighbor spacing distributions of domain centroids in the immiscible phase.
    %for quench time \( \tau_Q = 50 \), shown at \( t/t_{\rm eq} = 0.930 \), \( 0.967 \), and \( 1.0 \). 
    \updates{The quench time $\tau_{\Q}$ is fixed to $\tau_{\Q} = 50$. Here, $S$ is the distance scaled by the mean nearest-neighbor distance, and $t / t_{\rm eq}$ is the scaled time in units of the equilibration time $t_{\rm eq}$.}
    \updates{Panels (a--c) correspond to condensate component 
    %\( j=1 \) and \( j=2 \)
    1 of the binary Bose-Bose mixture}. Solid red lines represent the theoretical distribution from Eq.~
    %7 
    \updates{6}
    (main text) with \( k=2 \). \updates{Similarly, panels (d--f) correspond to the condensate component 2, with the red solid lines representing the theoretical prediction given in the Eq.~ 6 (main text) for \( k=2 \)}. Histograms are constructed from 
    %$\mathcal{R}=400$ 
    \updates{400}
    noise realizations with 25 bins.}
    \label{centroid_spacing}
\end{figure*}
For domain centroids, the nearest-neighbor spacing of component \( j=2 \) follows Eq.~
%7 
\updates{6 }
(main text) with \( k=2 \), as shown in Fig.~\ref{ppp-wigner}(d) at \( t/t_{\rm eq} = 0.930 \). 
In practice, the domain centroid $(\bar x, \bar y)$ is computed by considering equal weightage of each and every grid point constituting the domain (from binary density). We take the average of the coordinates within the binary domain given by $\bar x = \sum_{i=1}^{N} x_i/N$ and $\bar y = \sum_{i=1}^{N} y_i/N$, where $N$ is the number of grid points in a particular domain. The schematic representation of a domain and its geometric centroid is shown in the Fig.~\ref{centroids_schematic}.

We also compute the distributions of first-nearest-neighbor spacings at different times around the equilibration time \( t_{\rm eq} \). For vortices, these distributions are shown in Fig.~\ref{vortex_spacing}, with the upper panels corresponding to component \( j=1 \) and the lower panels to \( j=2 \). The corresponding distributions for domain centroids (mean domain positions) are presented in Fig.~\ref{centroid_spacing} with the same arrangement.

As discussed in the main text, the deviation of the first-nearest-neighbor spacing distribution from our KZM-PPP model as time goes on is expected due to the existence of the inter-species interaction coefficient $g_{12}$, affecting two different components ($j = 1$ and $j = 2$) each other and thus breaking ``randomly and independently distributed'' assumption of the homogeneous PPP for defects in each component. 
The interaction of defects in the same component also 
contributes to 
the deviation from our KZM-PPP model.

% As requested by the editor: 1) To avoid confusion please label the references in the Supplementary Information as ‘Supplementary References’
%\renewcommand\refname{Supplementary References}

\end{document}